\documentclass[preprint,11pt,authoryear]{elsarticle}
\usepackage{geometry}
\usepackage{amsthm}
\usepackage{hyperref}
\usepackage{algorithm}% http://ctan.org/pkg/algorithm
\usepackage[algo2e]{algorithm2e} 
\usepackage{algpseudocode}% http://ctan.org/pkg/algorithmicx
\setcitestyle{numbers,square}
\usepackage{graphicx}
\usepackage{url}
\usepackage{hyperref}
\usepackage{calligra}
\usepackage{optidef}
\usepackage{graphicx}
\usepackage{amsxtra}
\usepackage{amstext}
\usepackage{amssymb}
\usepackage{amsmath}
\usepackage{graphicx}
\usepackage{enumerate}
\usepackage{setspace}
\usepackage{color}
\usepackage{booktabs}
\usepackage{multirow}
\usepackage{float}
\usepackage{amsthm}
\usepackage{mathtools}
\usepackage{enumitem}
\usepackage{xr}

\newlist{UR}{enumerate}{3}
\setlist[UR]{label=Exp\arabic*:, leftmargin=1cm}

\DeclarePairedDelimiter\ceil{\lceil}{\rceil}
\DeclarePairedDelimiter\floor{\lfloor}{\rfloor}
\theoremstyle{definition}
\newtheorem{definition}{Definition}

\usepackage{subfig} % for subfigures
\usepackage{caption}

\newcount\Comments  % 1 suppresses notes to selves in text
\definecolor{darkgreen}{rgb}{0,0.5,0}
\definecolor{purple}{rgb}{1,0,1}
\definecolor{gold}{rgb}{0.71,0.58,0.06}

\newcommand{\edit}[1] {{\color{black}{#1}}}

\journal{Transportation Research Part C}

\begin{document}

\begin{frontmatter}



\title{\large \textbf{Automatic Vehicle Trajectory Data Reconstruction at Scale}}

\author[inst1,inst2]{Yanbing Wang}
\ead{yanbing.wang@vanderbilt.edu}
\author[inst2,inst3]{Derek Gloudemans}
\author[inst1,inst2]{Junyi Ji}
\author[inst3]{Zi Nean Teoh}
\author[inst4]{Lisa Liu}
\author[inst1,inst2]{Gergely Zachár}
\author[inst2]{William Barbour}
\author[inst1,inst2,inst3]{Daniel Work}

\affiliation[inst1]{organization={Department of Civil and Environmental Engineering, Vanderbilt University},country={United States}}
\affiliation[inst2]{organization={Institute for Software Integrated Systems, Vanderbilt University},country={United States}}
\affiliation[inst3]{organization={Department of Computer Science, Vanderbilt University},country={United States}}
\affiliation[inst4]{organization={Department of Electrical Engineering, Vanderbilt University},country={United States}}
\begin{abstract}
\edit{In this paper we propose an automatic trajectory data reconciliation to correct common errors in vision-based vehicle trajectory data. Given ``raw'' vehicle detection and tracking information from automatic video processing algorithms, we propose a pipeline including \textbf{(a) an online data association algorithm to match} fragments that describe the same object (vehicle), which is formulated as a min-cost network circulation problem of a graph, and \textbf{(b) a one-step trajectory rectification procedure formulated as a quadratic program} to enhance raw detection data. The pipeline leverages vehicle dynamics and physical constraints to associate tracked objects when they become fragmented, remove measurement noises and outliers and impute missing data due to fragmentations. We assess the capability of the proposed two-step pipeline to reconstruct three benchmarking datasets: (1) a microsimulation dataset that is artificially downgraded to replicate upstream errors, (2) a 15-min NGSIM data that is manually perturbed, and (3) tracking data consists of 3 scenes from collections of video data recorded from 16-17 cameras on a section of the I-24 MOTION system, and compare with the corresponding manually-labeled ground truth vehicle bounding boxes. All of the experiments show that the reconciled trajectories improve the accuracy on all the tested input data for a wide range of measures. Lastly, we show the design of a software architecture that is currently deployed on the full-scale I-24 MOTION system consisting of 276 cameras that covers 4.2 miles of I-24. We demonstrate the scalability of the proposed reconciliation pipeline to process high-volume data on a daily basis.}

\end{abstract}


% \begin{highlights}
% \end{highlights}

\begin{keyword}
trajectory data \sep 
data association \sep 
data reconciliation \sep
\end{keyword}

\end{frontmatter}
% \linenumbers

\section{Introduction}
\subsection{Motivation}

% Moreover, being able to generate traffic measurements live is also important to facilitate traffic control. Real-time traffic control strategies such as variable speed limit and ramp metering~\cite{carlson2010optimal}, or Lagrangian control with connected and autonomous vehicles~\cite{vinisky2018} require sufficiently prompt feedback and informative measurements to make decisions. To accomplish real-time deployment, we aim to build an efficient data reconciliation pipeline to handle streaming data with low latency.

% Ultimately, vehicle and transportation technologies are trending towards ``intelligent" and more autonomous solutions. Being able to measure the broader impact on traffic will be crucial in enabling future generations of traffic control and management. Now with advancement in camera resolutions and computer-vision algorithms for object detection and tracking, the process of getting reliable trajectory data can be greatly automated, yet issues on the quality of the data still remain. 

Vehicle trajectory data has received increasing research attention over the past decades. We draw particular attention to extracting high-resolution vehicle trajectory data from video cameras as traffic monitoring cameras are becoming increasingly ubiquitous. Video-based trajectory data collection in conjunction with image processing algorithms can offer a ``bird's eye'' perspective and, with a dense deployment of cameras, can achieve complete spatial and temporal coverage of a roadway segment. \edit{The capability to fully observe every vehicle for a long duration of time and over a long corridor is critical to advance research in areas such as training and calibration of traffic flow models, and driving behavior modeling~\cite{LI2020trajectory}.}

% Traditional traffic data collection mechanisms such as loop detectors provide low-frequency, aggregated features suitable for learning macroscopic characteristics; microscopic features, however, can be difficult to validate without the data at individual vehicle level. Vehicle on-board sensors and GPS measurements do not provide the necessary fidelity and/or completeness necessary for many applications such as vehicle energy/emission estimation~\cite{LI2020trajectory,coifman2017critical}. 

However, vision-based trajectory data by itself is of limited utility for traffic research because of noise and systematic sensing errors, thus necessitates proper processing to ensure data quality. Inaccuracies in such data -- projection errors, occlusions, dimensions, and kinematic physicality~\cite{PUNZO20111243}, for example -- need to be addressed. Automatic data reconciliation methods are needed to deal with massive volumes of trajectory data from traffic video processing.

\begin{figure}[htp]
    \centering
    \includegraphics[width=\linewidth]{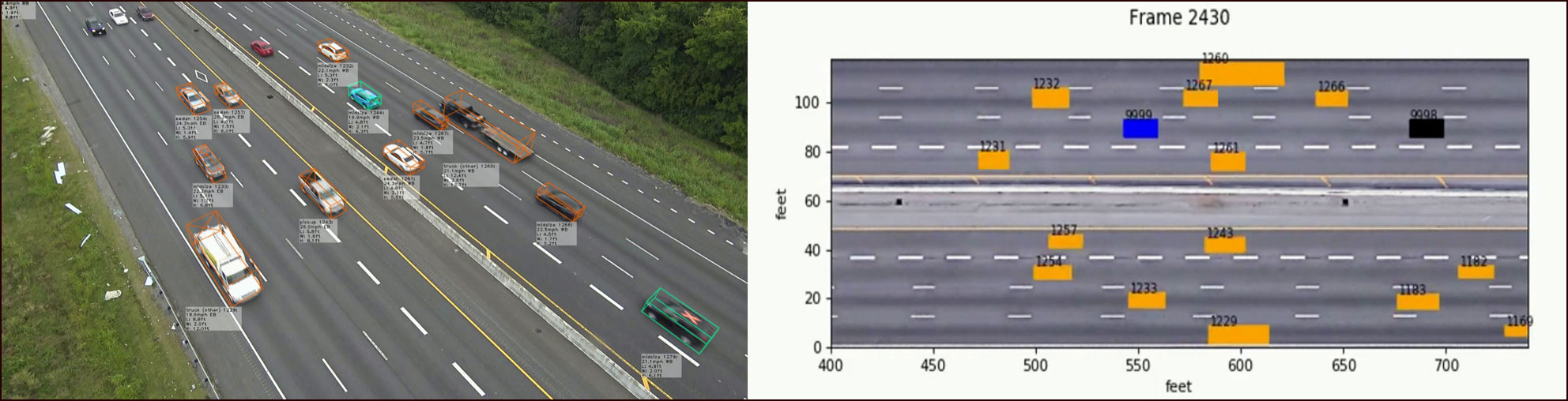}
    \caption{Left: the field of view from one of the 276 cameras with vehicles detected in 3D bounding boxes. Right: Projected positional measurements in roadway coordinates. }
    \label{fig:motion_cv}
\end{figure}

\edit{This paper describes the development of a two-step data reconciliation and postprocessing pipeline for a densely-instrumented camera network, I-24 MOTION~\cite{gloudemans202324}. This project was in collaboration with the Tennessee Department of Transportation (TDOT) and involved 276 fully-connected 4K cameras to monitor a 4.2-mile stretch of Interstate 24. The upstream process is an anonymous trajectory data generation pipeline using computer vision and image processing algorithms (detailed in~\cite{gloudemans202324, gloudemans2023interstate, gloudemans2023so}) to produce 3D vehicle detection and tracking data in a roadway reference system (Figure~\ref{fig:motion_cv}). The continuous field of view provides for complete, high-resolution trajectories for nearly every individual vehicle on the roadway.

A preliminary investigation on the raw trajectories revealed various errors arise from a combination of factors in upstream system design and computer vision detection and tracking, such as object hand-off across cameras, homography estimation, and the coordinate transformation from imagery to roadway coordinates. We have intentionally designed the post-processing step to be independent of the tracking algorithm and the specific error structures it may contain. 
 
 }
% Data quality and reliability from video-based extraction is not well-validated and no methods exist for rectifying this trajectory data in a rigorous and time-efficient manner. 
% \Yanbing{update the figures of detection errors}

% Vehicle trajectory reconstruction aims at correcting the upstream detection and tracking errors without re-processing the videos, but instead leveraging physics-based knowledge, such as vehicle motions, from a transformed roadway coordinate system. 

\subsection{Challenges}

\begin{figure}
    \centering
    \includegraphics[width=\columnwidth]{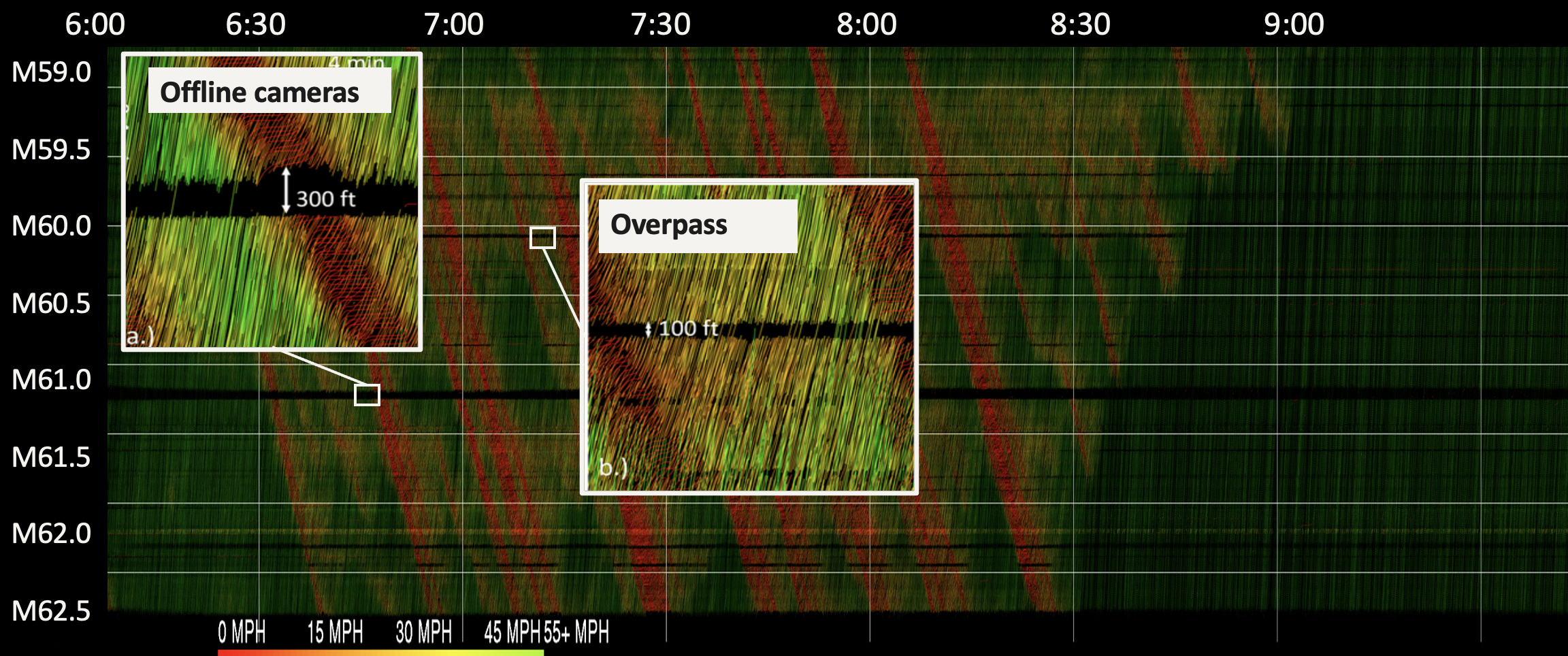}
    \caption{Time-space diagram for four hours of I-24 westbound morning rush hour traffic on Nov 22, 2022, generated from I-24 MOTION vehicle trajectories. x-axis: time of day (HH:MM); y-axis roadway postmile (mi). Fragments due to upstream artifacts are highlighted.}
    \label{fig:ts_fragments}
\end{figure}

\begin{figure}[H]
    \centering
    \includegraphics[width=\columnwidth]{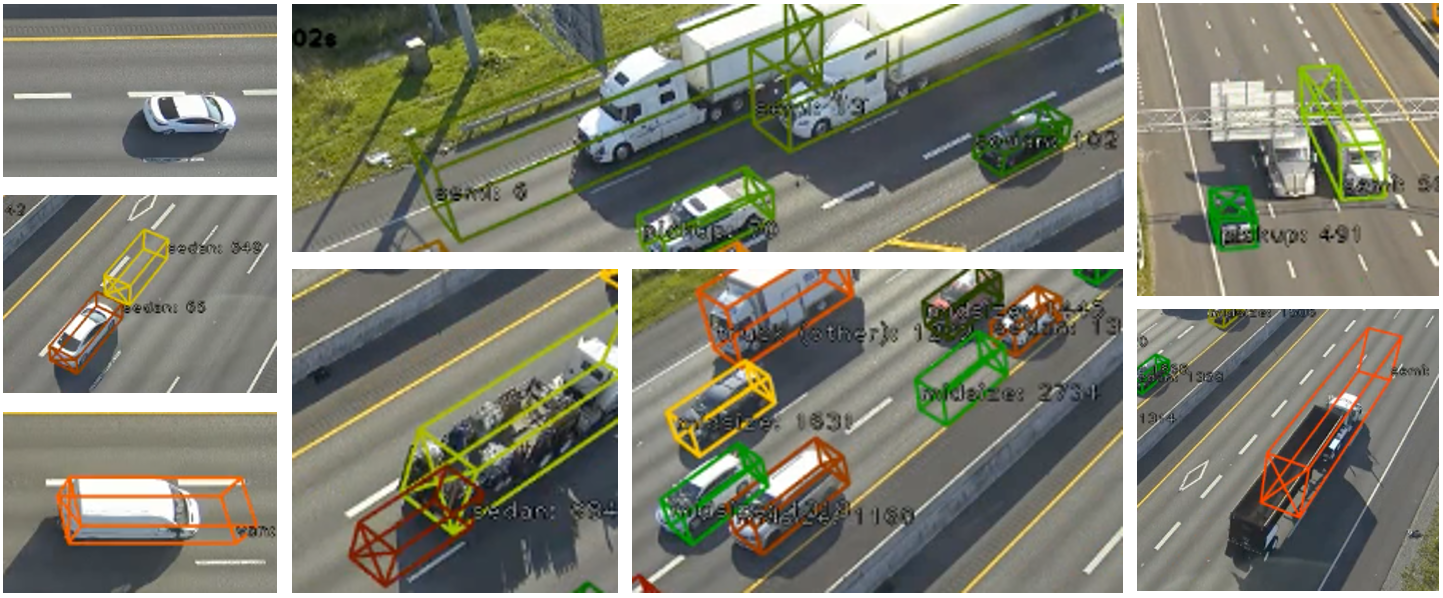}
    \caption{Examples of common inaccuracies from upstream video processing algorithms: missing detection, inaccurate localization, and false positives.}
    \label{fig:detection_errors}
\end{figure}

\edit{The first challenge in generating high-quality trajectory data through post-processing is the need to tackle a range of issues stemming from both upstream system design and computer-vision algorithms. One of the foremost challenges is dealing with ``tracking fragmentations", where the tracking of vehicles gets frequently disrupted, often due to errors in tracking algorithms, object occlusions, inaccuracies in the transformation from image space to roadway coordinates, and data segmentation due to distributed computation (see Figure~\ref{fig:ts_fragments}). Additional issues include missing detections, false positives, tracking inconsistencies, and inaccurate localization (see Figure~\ref{fig:detection_errors}). These errors are common in video-processing algorithms, which necessitate the significance of having a comprehensive data post-processing procedure capable of addressing a broad spectrum of issues before the trajectory data can be employed for traffic analysis purposes.

A secondary challenge is that the data postprocessing algorithms should be efficient in order to handle the growing volume of data coming from large-scale real-world camera systems that are designed to enable real-time traffic monitoring and control. Data reconciliation must be performed in real-time or with low latency to facilitate continuous or even live generation of high-volume data. Therefore, fast and online data reconciliation methods are to be explored.

Last but not the least, while many prior studies in the literature have acknowledged the importance of data cleaning and have proposed various methodologies, there is a noticeable absence of open-sourced datasets and implementations suitable for method benchmarking. A reliable benchmarking dataset should contain raw tracking results and their corresponding ground truth. To initiate benchmarking endeavors, we are making our datasets available (both simulated data and raw tracking data with manually labeled ground truth), along with the post-processing repository used in this paper, accessible to the research community.
}

\subsection{Problem statement and contributions}
\label{sec:problem}
% Motivated by the lack of a high-quality, large-scale source of trajectory data necessary to assess the impact of driving behavior on large-scale traffic, 

\edit{This paper focuses on the postprocessing component as part of the I-24 MOTION software system. The postprocessing takes the raw tracking information from upstream computer vision pipeline as input, and outputs trajectory data that is more complete and with kinematic noises and outliers removed. The postprocessing pipeline is designed to be agnostic of the specific upstream system design and tracking configurations, e.g., setup of camera clusters, image-to-roadway transformation and object hand-off between cameras. All of these design decisions have been detailed in~\cite{gloudemans2023interstate,gloudemans202324,gloudemans2023so}. Instead, the postprocessing pipeline addresses two common tracking problems: the first is \textit{fragmentation}, meaning the tracking of a vehicle is interrupted constantly, resulting in multiple short, partially-tracked fragments; the second is the presence of \textit{noises, outliers and missing data} on each trajectory (Figure~\ref{fig:detection_errors}), due to miscellaneous detection and tracking errors upstream. The raw tracking input is in a 2D relative roadway coordinate system (a locally perpendicular coordinate system with the primary axis being the direction of travel and the secondary axis perpendicular to the direction of travel). The postprocessed output trajectories are in the same coordinate system.}

We highlight the following contributions in this article:

\edit{
\begin{itemize}
% \item We propose an automatic trajectory data reconciliation pipeline for a modern traffic testbed. Specifically, the pipeline includes
% a) an online data association algorithm to solve a min-cost flow problem, which consequently matches fragments that belong to the same object, and
% b) a novel trajectory reconciliation algorithm, which is formulated as a quadratic program. It reconstructs realistic vehicle dynamics from disturbed detection data with systematic smoothing and outlier correction. The resulting trajectories automatically satisfy the internal consistency (differentiation of trajectories with speeds and accelerations).

    \item To solve the \textit{fragmentation} issue, we develop an efficient online data association algorithm, \textit{online negative cycle canceling}, which overcomes the limitations of batch methods for solving the data association problem as a network flow problem. The online algorithm ensures optimality of the batch algorithm, and is able to process up to half a million fragments per hour as required by the streaming data rate from the upstream tracking system. Benchmark experiments show that the proposed data association algorithm has the capability to combine together 60-77\% of fragments, resulting in trajectories that are 3-4.5 times longer.

    \item To address the \textit{noises, outliers and missing data}, we introduce an one-step convex program to rectify trajectories. This method is capable of simultaneous imputation of missing data, denoising of signals, and removal of outliers. This approach, based on \textit{elastic net regularization}, ensures kinematic feasibility and internal consistency of trajectories and is able to streamline the process compared to other multi-step methods.

    \item We design a software architecture for postprocessing that utilizes asynchronous processes and is currently deployed on the I-24 MOTION system to continuously generate a large volume of high-quality open-road trajectory data. The online nature of the data association algorithm and the convex trajectory rectification formulation allow this deployment to potentially work with streaming raw tracking data from upstream.

    \item We release the datasets used to benchmark the proposed postprocessing algorithms in this paper, including the ground truth, manually-perturbed and reconciled simulated data, the raw tracking data from three scenes and the corresponding manually labeled ground truth and reconciled data. Additionally, we provide the postprocessing repository used in our paper for the community to start the benchmarking efforts in the direction of trajectory data reconciliation [link removed for review purposes].
% \item We provide evaluations on the trajectory quality including dynamics analysis and error statistics. Results show that the reconciled trajectories improve a variety of measures on various traffic scenarios, regardless of upstream data source (simulation data or raw tracking data from real video processing algorithms).
\end{itemize}
}

% This work, although still in progress, illustrates the first step towards automatic and online trajectory data reconciliation from video-based data extraction.  Future work will focus on fine tuning the parameters associated with the proposed algorithms to further enhance the data quality. 
% This work aims to provide an efficient pipeline to systematically rectify trajectory data produced by automatic computer vision algorithms. \edit{, and it generates all trajectories that are kinematically feasible, i.e., the velocities and accelerations fall within feasible ranges.} The work presented here is preliminary, but outlines the core methods needed for trajectory cleaning. Detailed algorithm tuning will increase the overall accuracy of the methods prior to deployment on a large scale system on a freeway testbed.

\edit{The rest of the paper is organized as follows: Section~\ref{sec:related_work} summarizes available vehicle trajectory datasets, data acquisition methods and their scopes. Section~\ref{sec:problem_formulation} specifies the problem formulation for solving the fragmentation and trajectory rectification problems. In Section~\ref{sec:experiments}, we assess the capability of the proposed pipeline to reconstruct three benchmarking datasets: (1) a microsimulation dataset that is artificially downgraded to replicate upstream errors, (2) a 15-min NGSIM data that is manually tampered, and (3) tracking data consists of 3 scenes from a subsection of the I-24 MOTION system. All of the experiments show that the postprocessed trajectories improve the accuracy on all the tested input data for a wide range of measures. Additionally we demonstrate the design of a software architecture that is capable of processing high-volume data from the full-scale I-24 MOTION system. Finally in Section~\ref{sec:conclusions}, we summarize the methods, results and highlight potential future improvements.
}

\section{Related work}
\label{sec:related_work}
\subsection{Vehicle trajectory datasets}
Vehicle trajectory data is integral to the study of traffic dynamics. They reveal the relationship between individual traffic participants and the resulting traffic flow phenomena. The increasing volume of data sources is enabling revolutionary research on, for example, traffic flow theory~\cite{treiterer1974hysteresis,hoogendoorn2013traffic,greenberg1959analysis} and traffic state estimation~\cite{seo2017traffic,wang2005real,tao2012real}.
Trajectory data contains the microscopic fidelity (fine-grained vehicle positions) necessary for building and validating realistic microscopic models. Accurate estimation of energy consumption also relies on vehicle-level detailed dynamics~\cite{fiori2016power,fiori2020energy,oh2020vehicle}. \edit{The accurate quantification of the energy demand of vehicles is of critical importance for many applications, including the design of modern energy-centric intelligent transportation
systems. Available high fidelity energy modeling tools such as Autonomie~\cite{halbach2010model}, CarSim and TruckSim~\cite{mechanical2014carsim} relies on realistic acceleration and velocity dynamics for validation.} High-quality trajectory data can close the gap for understanding microscopic traffic phenomena, such as lane-change and car-following~\cite{gazis1959car,gipps1981behavioural,schakel2012integrated,kesting2008calibrating,treiber2013traffic} and the impact of mixed autonomy in traffic~\cite{stern2018dissipation,gunter2019are,yang2019impacts}.

Small-scale trajectory data is usually collected via vehicles that have been equipped with high-accuracy GPS sensors~\cite{jiang2014traffic, JIANG2015338, coifman2016collecting,huang2018experimental} and other in-vehicle sensors. Unlike conventional probe vehicles or floating cars, vehicles currently collect their position data with significantly higher spatial and temporal resolution levels to obtain high-quality trajectory data~\cite{guo2019urban}. Data collected from a combination of in-vehicle sensors such as radars, LiDAR and cameras provides ambient information, which can be useful for traffic flow studies as well as for training and testing AI algorithms for autonomous driving. Readers are directed to references such as~\cite{makridis2020openacc, kang2019test} for a comprehensive discussion on such dataset. 
% Throughout the literature, we have found a growing volume of trajectory datasets collected at various scales through different sensing mechanisms. We summarize these in Table~\ref{tab:data_comparison}. 

Image sensors and video processing algorithms associated with overhead cameras or Unmanned Aerial Systems (UAV) have become increasingly powerful as a large-scale trajectory data source. The seminal work is the Next Generation Simulation (NGSIM) dataset~\cite{NGSIM2006} developed in early 2000s, which is a collection of real-world trajectories, based on the use of cameras mounted on tall buildings and covering approximately 600-meter long roadway sections with a frequency of 10 Hz in several US locations. More recently in the highD dataset~\cite{krajewski2018highd}, videos were recorded by overhead flying drones (4096 × 2160 pixels, 25 fps), which provided much higher-quality videos than that of the NGSIM dataset~\cite{coifman2017critical}. HighD provides highway trajectories covering 420m of road segment, for a total of 28,000 vehicle miles traveled. The pNEUMA large-scale field experiment~\cite{barmpounakis2020new} recorded traffic streams in a multi-modal congested environment over an urban area using UAV. The dataset was generated by a swarm of 10 drones hovering over a traffic intensive area of 1.3 km$^2$ in the city center of Athens, Greece, covering more than 100 km-lanes of road network at 25Hz. \edit{Zen Traffic Data~\cite{seo2020evaluation} is obtained by cameras installed on some of the light poles of the Hanshin Expressway to observe all vehicles in target sections at a 0.1 second interval in Osaka, Japan}. More recently, Automatum dataset became available~\cite{spannaus2021automatum}, which covers 12 characteristic highway-like scenes from 30 hours of drone videos. HIGH-SIM~\cite{Shi2021} is another high-quality highway trajectory dataset extracted from aerial video data. The videos were collected by three 8K cameras in a helicopter over an 8000 ft long segment of the I-75 freeway in Florida for 2 hours. The dataset covers a wide range of traffic characteristics. A comparison across selected datasets can be viewed from Table~\ref{tab:data_comparison}. Other examples and discussion on their usage in traffic studies can be found in a comprehensive survey, such as~\cite{LI2020trajectory}. 

% \usepackage{booktabs}
\begin{table}
\centering
\tabcolsep=0.11cm
\begin{tabular}{lllllllll}
% \begin{tabular*}{\textwidth}{l @{\extracolsep{\fill}}lllllll}
\toprule
Dataset     & Year & Context        & \begin{tabular}[c]{@{}l@{}}Camera \\ config\end{tabular} & \begin{tabular}[c]{@{}l@{}}Road \\ segment\end{tabular}    & VMT & \begin{tabular}[l]{@{}l@{}}Hours of \\ recording\end{tabular}    & \\ \midrule
NGSIM~\cite{NGSIM2006}       &2006 & Highway        & 8 cameras        & 600 m         & 18,000           & 0.75    \\
pNeuma~\cite{barmpounakis2020new}      &2018 & Urban arterial & 10 drones & 10 km         &   N/A      & 59             \\
highD~\cite{krajewski2018highd}       &2018 & Highway        & 1 drone    & 420 m         & 28,000         & 147    \\
\edit{Zen Traffic}~\cite{seo2020evaluation} & \edit{2020} & \edit{Highway} &
6 cameras& \edit{2 km} & \edit{N/A}& \edit{5}\\
Automatum~\cite{spannaus2021automatum}    &2021 & Highway        &  \begin{tabular}[c]{@{}l@{}}Overhead \\ drones\end{tabular}  & N/A  & 18,724              & 30     \\
HIGH-SIM~\cite{Shi2021}    &2021 & Highway        &  \begin{tabular}[c]{@{}l@{}}Aerial \\ cameras\end{tabular}  &  2,438 m  &    N/A           &     2   \\
\begin{tabular}[c]{@{}l@{}}I-24 \\ MOTION~\cite{gloudemans202324}\end{tabular} & 2022 & Highway        & \begin{tabular}[c]{@{}l@{}}276 \\ 4K cameras\end{tabular}     & 4.2 miles      &  \begin{tabular}[c]{@{}l@{}}200M/yr\\ (expected) \end{tabular}       & daylight    \\ \bottomrule
\end{tabular}
\caption{Video-based trajectory data comparison. VMT: vehicle miles traveled}
\label{tab:data_comparison}
\end{table}

\subsection{Data extraction and quality assessment}
% \Yanbing{cite makdris from ETH 2023}
Raw trajectories from video footage are usually obtained from a \textit{tracking-by-detection} framework~\cite{kalal2012tracking,Andriluka2008people}, where a set of object detections for all frames are linked across time to form continuous trajectories. The detected bounding boxes are then transformed to a common reference system for meaningful interpretation of dynamics. One of the errors we aim to reconcile is tracking discontinuity due to, for example, noisy detection, inevitable object occlusion and switching camera field of views, which results in track fragmentations. The first problem to address is data association, with is given a set of detections, identify their origin (objects that they are associated with). Data association problems can be difficult when the tracking produces large fragmentations or when the detections have false alarms or missing detections, which are common in real-world video-based detection systems. 

\edit{Different approaches for data association have been proposed depending on the association criteria, complexity of object motion and the computation requirements. In general, these approaches differ from their choices of 1) matching cost (sometimes termed as probability, affinity, energy or confidence) and 2) matching criteria (Hungarian algorithm~\cite{perera2006multi} and bipartite graph matching~\cite{bewley2016simple, wojke2017simple} etc.), which result in various problem formulations. The matching cost reflects kinematic~\cite{dicle2013way,yoon2015bayesian} and/or appearance information~\cite{bewley2016alextrac, kim2015multiple, bae2014robust, geiger20133d}, and the matching criteria directly drives the algorithm to solve the data association problem. }

Graph-based formulations provide efficient algorithms for finding the global min-cost tracking solutions. The tracks (or detections) can be naturally represented as the nodes of a graph and the pair-wise matching costs can be represented as graph edges. The general data association problem is finding the least-cost set cover on the track graph~\cite{chong2012graph}. Studies such as~\cite{castnnon2011multi, zhang2008global,vyahhi2008tracking} have explored efficient algorithms related to bipartite matching and min-cost flow. Interested readers are referred to a recent survey, such as~\cite{Rakai2022} on this topic. 

% Finding such best hypothesis is an NP-hard matching problem and requires combinatorial optimization algorithms. However, special properties of the graph, such as Markov assumption of the association cost, can be utilized so that polynomial-time algorithms (e.g., bipartite matching and min-cost flow solvers) can be applied~\cite{castnnon2011multi, zhang2008global,vyahhi2008tracking}.While the subject of data association or multi-object-tracking (MOT) has a long history in computer vision research, a complete review lies beyond the scope of this paper. Interested readers are referred to a recent survey, such as \cite{Rakai2022} on this topic. 

Although there has been a growing projects dedicated to generate large vehicle trajectory data, the quality issues remain insufficiently addressed. Many works~\cite{coifman2017critical,PUNZO20111243,thiemann2008estimating} raised the issue of the quality of the seminal NGSIM data, including nonphysical kinematics and inter-vehicle distance. The importance for data postprocessing is also emphasized in~\cite{kesting2008calibrating,PUNZO20111243,treiber2013traffic,LI2020trajectory}. Montanino \& Punzo~\cite{montanino2015trajectory} undertook one of the most thorough efforts of data reconciliation by applying a series of smoothing operations on the dataset. They posed the smoothing and reconstruction as a nonlinear, non-convex optimization problem, as to find the minimum local smoothing window subject to kinematic constraints. Other data postprocessing efforts~\cite{thiemann2008estimating, hamdar2008driver,duret2008estimating} also performed data reconciliation considering realistic vehicle kinematics and driving dynamics. \edit{Although with the general goal of improving overall quality of the trajectory data, these methods provide treatments for only limited types of errors, as summarized in Table~\ref{tab:method_comparison}. Most of the algorithms do not have capabilities to operate online or provide provable optimality, which prevent them from being deployed to a large-scale system with raw tracking data as streaming input. There are many practical challenges to benchmark these trajectory reconstruction methods. For example, we have not yet found a valid benchmark dataset that consists of both raw tracking results and the corresponding ground truth. In this work we demonstrate a more efficient alternative for trajectory reconciliation that is capable to treat a wide range of errors commonly seen in video processing algorithms, and process high volume of streaming data. We also release the datasets (both the simulated data and the raw tracking data with manually labeled ground truth) and the postprocessing repository used in our paper for the community to start the benchmarking efforts in this direction.}

% \usepackage{booktabs}
\begin{table}[]
\centering
\tabcolsep=0.11cm
\begin{tabular}{@{}llllllll@{}}
\toprule
 Reference    & \multicolumn{5}{c}{Treatments}                                                                                                                                                                                & \multicolumn{2}{c}{Algorithm properties}                                            \\ 
\cmidrule(lr){2-6} \cmidrule(lr){7-8}
& \begin{tabular}[c]{@{}l@{}}Data \\ association\end{tabular} & Denoise & \begin{tabular}[c]{@{}l@{}}Data \\ imputation\end{tabular} & \begin{tabular}[c]{@{}l@{}}Outlier and \\ false positives\end{tabular} & \begin{tabular}[c]{@{}l@{}}Conflict \\ resolution\end{tabular} & Online? & \begin{tabular}[c]{@{}l@{}}Provable \\ optimality?\end{tabular} \\
\midrule
Montanino \cite{montanino2013making}  &       &   \checkmark  &      &  \checkmark &         &   &     \\
Montanino \cite{montanino2015trajectory}  &       &   \checkmark  &     &  \checkmark &       \checkmark    &   &     \\
Coifman \cite{coifman2017critical}   &   &   \checkmark      &      &   \checkmark   &  \checkmark &  \multicolumn{2}{c}{Manual correction} \\
Punzo~\cite{PUNZO20111243}  &       &   \checkmark  &      &     \checkmark     &   \checkmark   &     &        \\
Thiemann~\cite{thiemann2008estimating}  &       &   \checkmark  &      &     &   &         &        \\
% Duret~\cite{duret2008estimating} &       &   \checkmark  &      &        &   &      &        \\
% Treiber~\cite{treiber2013traffic}&       &   \checkmark  &      &        &    &     &        \\
% Kesting~\cite{kesting2008calibrating}&       &   \checkmark  &      &        &   &      &        \\
Hamdar~\cite{hamdar2008driver} &       &   \checkmark  &      &     &   &         &        \\
Fard~\cite{fard2017new}  &       &   \checkmark  &      &  \checkmark   &   &         &        \\
Ji~\cite{ji2023sdr} &       &   \checkmark  &   \checkmark    &    & \checkmark   &     &     \checkmark   \\
\textbf{Ours}      &   \checkmark       &   \checkmark  &      \checkmark  &     \checkmark                                             &     &      \checkmark     &  \checkmark   \\ \bottomrule
\end{tabular}
\caption{A comparison on trajectory data reconstruction methods.}
\label{tab:method_comparison}
\end{table}

%~\cite{thiemann2008estimating, hamdar2008driver,duret2008estimating}
% ~\cite{coifman2017critical,PUNZO20111243,thiemann2008estimating} 
% ~\cite{kesting2008calibrating,PUNZO20111243,treiber2013traffic,LI2020trajectory}.

\section{Problem formulation}
\label{sec:problem_formulation}
We focus on addressing two types of issues stemming from the raw tracking data. \edit{The first is a \textit{fragmentation}, in which an object (vehicle) is tracked multiple times, resulting in partially tracked short trajectories (or fragments). Fragments occur because tracking is disrupted due to occlusion by larger vehicles or bridges, or the vehicles move across cameras or computational boundaries} (Figure~\ref{fig:ts_fragments}). It results in discontinuity in tracking and broken trajectories. The second type of errors include \textit{noises, outliers and missing data} which are caused by either a missing detection, errors in vehicle position estimation, or false positives produced by the object detector (shown in Figure~\ref{fig:detection_errors}). Given these known errors, we propose a two-step trajectory reconstruction approach: \edit{1. an \textit{online negative cycle cancellation} algorithm to solve the fragment/data association problem formulated as a network flow problem, and 2. a \textit{convex program} to rectify trajectories such that their high-order derivatives are kinematically feasible (i.e., velocities and accelerations are within a feasible range), and the gaps in between associated fragments are filled in and outliers are removed.}

%\derek{This is a bit of a tough claim. It's gonna be hard for a traffic community person to understand how the detection-originated errors are significantly different from the tracker-based errors.}\Yanbing{updated} 

% behavior of the upstream tracker. This error leaves the majority of the trajectories to be fragmented, especially when an object (vehicle) transitions through one camera field of view to another, or when a vehicle is occluded by a larger vehicle for a sustained period. This error cannot be ignored, because it causes double-counting of vehicle numbers and infers incorrect traffic flow and density information. Furthermore, short fragments are not informative for studying, for example, driving behavior, which benefits from longer car-following data. The second source of error is from detection (Figure~\ref{fig:detection_errors}), which produces missing detection, wrong localization, non-rectilinear shapes etc. These errors are observed to be sparse, and can be rectified by enforcing physical constraints, such as fixed vehicle dimension and smoothness in kinematics. 

The methods proposed below are performed on the 2D raw trajectories in a relative roadway coordinates, assuming a fixed point on each vehicle travels in 2D planar motion. It follows a local coordinate system such that the $x$-axis is parallel to the lane lines, and the $y$-axis is perpendicular to $x$.

In this section, we outline the problem formulation for multi-object tracking as a \edit{minimum-cost circulation (MCC)} of a graph. Solving for MCC on a track graph results in trajectory sets that have the highest \textit{maximum a posteriori} (MAP) probability. The problem formulation is explained in literature such as~\cite{zhang2008global, wang2019mussp, wang2020efficient, wang2023onlinemcf}, and therefore only highlighted briefly in this section.

\subsection{Preliminary: MCC and negative cycle cancellation}

A fragment with index $k$ is denoted as $\phi_k = \{p_1,...,p_n\}$, which consists of a series of positional data ordered by time (frame). Each data point $p_i$ is a vector containing timestamp, $x$ and $y$ position of a fixed point on the bounding box. We are given a set of fragments as input $\Phi = \{\phi_i\}$. A trajectory $\tau_k=\{\phi_{k_1},...,\phi_{k_n}\}$ consists of one or multiple fragments. A set of such trajectories form a trajectory set hypothesis $T = \{\tau_1, ..., \tau_K\}$. Assuming that fragments are \edit{conditionally independent}, the fragment association step aims at finding $T^*$, the hypothesis with the highest MAP:

\begin{equation}
\label{eq: MAP}
    \begin{aligned}
    T^* &=\textrm{\edit{argmax}}_{T}P(T|\Phi)\\
        &=\textrm{\edit{argmax}}_{T}P(\Phi|T)P(T)\\
        &=\textrm{\edit{argmax}}_{T}\prod_iP(\phi_i|T)\prod_{\tau_k\in T}P(\tau_k)\\
        & \textrm{s.t. }  \tau_k \cap \tau_l = \emptyset, \ \forall k \neq l,
    \end{aligned}
\end{equation}
with a non-overlapping trajectory constraint, since each fragment can belong to at most one trajectory. The likelihood $P(\phi_i|T) = P(\phi_i) = \beta_i$ indicates the probability that fragment $i$ is a false positive and thus should not be included in the trajectory hypothesis. The prior of a trajectory can be modeled as a Markov chain:

\begin{equation}
    P(\tau_k) = P_{enter}(\phi_{k_1})\prod_{i=1}^{n-1}P(\phi_{k_{i+1}}|\phi_{k_{i}})P_{exit}(\phi_{k_n}),
\end{equation}
where $P_{enter}(\phi_{k_1})$ and $P_{exit}(\phi_{k_n})$ denote the probabilities that $\phi_{k_1}$ starts the trajectory and $\phi_{k_n}$ ends the trajectory, respectively. Taking the negative logarithm of~\eqref{eq: MAP}, the MAP problem becomes equivalent to the following integer program:

\begin{subequations}
\label{eq: ilp}
    \begin{align}
    \underset{f_i, f_{i,j}, f_i^{en},f_i^{ex}}{\textrm{minimize}} \ & \sum_i c_if_i + \sum_i c_i^{en}f_i^{en} +\sum_{i,j} c_{i,j} f_{i,j}+\sum_i c_i^{ex}f_i^{ex} \\
    \textrm{s.t.} & \ f_i, f_{i,j}, f_i^{en},f_i^{ex}\in \{0,1\}, \  \label{unit_capacity} \\ 
     & \ f_i^{en} + \sum_{j} f_{j,i}  = f_i = f_i^{ex} + \sum_{j} f_{i,j},\ \forall i \in V \char`\\ \{s,t\},\label{conservation}
    \end{align}
\end{subequations}
where
\begin{equation}
     c_i^{en} = -\log P_{enter}(\phi_i), \ c_i^{ex} = -\log P_{exit}(\phi_i), \ c_{i,j}=-\log P(\phi_i|\phi_j), \ c_i = -\log \dfrac{1-\beta_i}{\beta_i}.
\end{equation}

The decision variables are binary according to constraint~\eqref{unit_capacity}. $f_i$ indicates whether $\phi_i$ should be included in any trajectory, $f_i^{en}$ and $f_i^{ex}$ determine whether a trajectory starts or ends with $\phi_i$, respectively. $f_{i,j}$ indicates if fragment $\phi_j$ is an immediate successor of $\phi_i$.
Constraint~\eqref{conservation} ensures non-overlapping trajectories. \edit{The situations when an object is tracked by multiple fragments are addressed automatically in the MCC formulation: the fragments that have overlapped time intervals will not be associated into the same trajectory. The most likely association hypothesis is determined by a combination of $\beta_i, c_i^{en}, c_i^{ex}$ and $c_{i,j}$.}

% Optimal solution to~\eqref{eq: min_cost_flow} is guaranteed according to~\cite{ahuja1988network}. 

% In seminal work~\cite{zhang2008global}, it is shown that~\eqref{eq: ilp} has a natural graph interpretation, and solving for~\eqref{eq: ilp} is equivalent to finding the min-cost-circulation on a tracklet graph, which has a polynomial solution~\cite{ford1956maximal}.
% The graph is constructed such that each fragment $\phi_i$ is represented as two nodes $u_i$ and $v_i$, with a directed edge $(u_i, v_i)$ and a cost indicating \textit{inclusion} of $\phi_i$; edges between two fragments $\phi_i$ and $\phi_j$ are represented as $(v_i, u_j)$, with the cost related to the likelihood of $\phi_j$ following $\phi_i$. The edge direction implies the sequential order between fragments. The resulting graph is therefore a directed acyclic graph (DAG), see Figure~\ref{fig:mot_graph}. Furthermore, the graph has
% a source node $s$ that has an incident edge to every $u$, and a sink node $t$ that every $v$ points to. We denote the weighted DAG as $G(V,E,C)$, with node set $V$, edge set $E$ and edge weights (costs) $C$. The data association problem can be formulated as finding a set of non-overlapping $s-t$ paths on a DAG with lowest total cost, or finding ``flows" from $s$ to $t$ that result in the minimum total cost. 

\begin{figure}[ht]
    \centering
    \includegraphics[width = \linewidth]{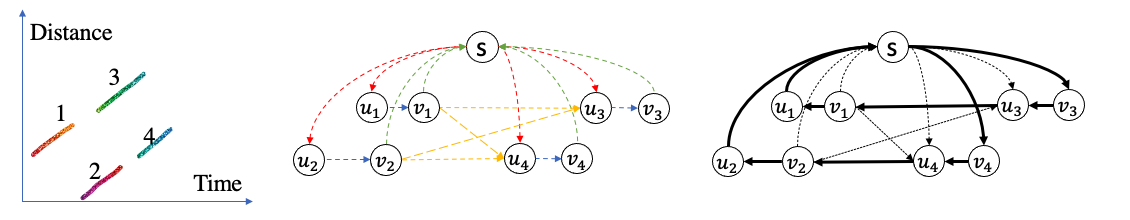}
    \caption{Left: fragments in time-space coordinates. In this example the correct association is \{$\phi_1$, $\phi_3$\} and \{$\phi_2$, $\phi_4$\}. The numbers indicate the order of last timestamp. Middle: fragments as a circulation graph. Red edges are the entering edges with cost $c_i^{en}$; blue edges are inclusion edges with cost $c_i$; green ones are exiting edges with cost $c_i^{ex}$ and yellow ones are transition edges with cost $c_{ij}$. Right: the residual graph after running the negative cycle canceling algorithm to obtain the min-cost circulation. The residual edges that carry the min-cost circulation are highlighted in bold. The fragment association assignment can be obtained by tracing along the bold edges.}
    \label{fig:mot_graph}
\end{figure}

In seminal work~\cite{zhang2008global}, it is shown that~\eqref{eq: ilp} has a natural graph interpretation, and solving for~\eqref{eq: ilp} is equivalent to solving the min-cost-flow of a tracklet graph, which has a polynomial solution~\cite{ford1956maximal}. Later in the work of~\cite{wang2020efficient}, it is proven that the min-cost-flow problem for MOT is equivalent to a min-cost-circulation problem on a slightly modified graph. Many efficient algorithms are developed to solve this problem~\cite{ahuja1988network,Sokkalingam2000polynomial,Klein1966APM,goldberg1989finding}. 
The graph is constructed such that each fragment $\phi_i$ is represented as two nodes $u_i$ and $v_i$, with a directed edge $(u_i \rightarrow v_i)$ and a cost $\$(u_i \rightarrow v_i)=c_i$ indicating \textit{inclusion} of $\phi_i$; edges between two fragments $\phi_i$ and $\phi_j$ are represented as $(v_i \rightarrow u_j)$, with the cost $\$(v_i \rightarrow u_j)=c_{ij}$ related to the likelihood of $\phi_j$ following $\phi_i$. The edge direction implies the sequential order between fragments. Furthermore, the graph has
a dummy node $s$ that has an incident edge to every $u$, and every $v$ directs back to $s$. The resulting graph is therefore a directed circulation graph, see Figure~\ref{fig:mot_graph}. We denote this circulation graph as $G(V,E)$, with node set $V$ and edge set $E$. Each edge $e:=(u,v) \in E$ has a unit capacity $r(e)=1$, a cost $\$(e)$ and a binary flow $f(e)\in \{0,1\}$. The data association problem can be formulated as finding a set of non-overlapping circulations $f$ on $G$ with the lowest total cost. The total cost of the circulations is $\sum_{e \in f} \$(e) f(e)$.

% The cost functions for the edges are described next. There are two types of edges, one type of edges stem from the source node $s$ or connect to the sink node $t$, which we term \textbf{enter/exit edges}. The cost on such edges are zero, i.e., there is no penalty or reward to start or end a trajectory from any fragment. The other type of edges are \textbf{transition edges} that connect one fragment to another. Costs are specified on these edges to drive to the optimal solution: paths from $s$ to $t$ that contribute to the mininum cost of the network, and are therefore the stitched trajectories (see Figure~\ref{fig:mot_graph} for an example). Details follow.

% \subsection{Negative cycle cancellation}

One efficient algorithm is the negative cycle canceling algorithm (NCC) proposed by Klein~\cite{Klein1966APM} and later on optimized by Goldberg et al.~\cite{goldberg1989finding,Sokkalingam2000polynomial}, based on the Ford-Fulkerson's method for incremental improvement. To understand the algorithm we first recall the definition of an important concept -- a residual graph $G_r$:

\begin{definition}
\label{def: residual_graph}
The residual graph $G_r(V, E_r)$ for the original directed graph $G(V,E)$ with respect to a flow $f$ is generated by replacing each edge $e=(u \rightarrow v)\in E$ by two residual edges $e^{\prime}=(u\rightarrow v) \in E_r$ and $e_r =(v\rightarrow u) \in E_r$, with cost $\$(e^{\prime})=\$(e) $ and residual capacity $r(e^{\prime})=r(e)-f(e)$, while $\$(e_r)=-\$(e) $ and $r(e_r)=f(e) $.
\end{definition}
In the context of MOT graph as shown in Figure~\ref{fig:mot_graph}, the construction of residual graph can be simplified. The edges in the flow of the original graph simply needs to be reversed and costs on the edges negated, to form the corresponding residual graph.

The idea of NCC is to repeatedly find a cycle with negative cost in the residual graph $G_r$ and push flow through the cycles. The algorithm terminates when no more negative cycles can be found (optimality condition). We direct interested readers the above reference for the details and proof of correctness of this algorithm, and only provide an outline in Algorithm~\ref{alg:ncc}. 

First, a circulation graph $G(V,E)$ is constructed from the set of fragments $\Phi$ (ConstructTrackletGraph) and we iteratively look for a negative cycle in $G_r$ based on, for example, Bellman-Ford algorithm. If such cycle exists, then update the residual graph according to Definition~\ref{def: residual_graph} (PushFlow). When the iteration stops (no more negative cycle can be found), the assignment, or the trajectories, can be extracted by traversing along all the cycles through the residual edges in $G_r$ (FlowToTrajectories). 

\begin{algorithm}
\caption{Negative cycle cancellation for min-cost-flow on a tracklet graph}
\label{alg:ncc}
\SetKwInput{KwData}{Input}
\SetKwInput{KwResult}{Result}
\SetAlgoLined
    \KwData{Set of fragments $\Phi=\{\phi_i\}$}
    \KwResult{Set of trajectories $T=\{\tau_i\}$}

\quad $G(V,E,C)\leftarrow$ ConstructCirculationGraph($\Phi$)

\quad $f \leftarrow 0$

\quad $G_r \leftarrow G$

% \quad $\pi^{(0)}, d^{(0)} \leftarrow$ ShortestPath($G_r^{(0)}$) //run single source shortest path from s. $\pi$ is the shortest path from s to t, $d$ is a distance (cost) map from s to any other nodes in $G$

\While {a negative-cost cycle $\Gamma$ exists in $G_r$}{
    // Update residual graph\\
    $G_r \leftarrow$ PushFlow($G_r, \Gamma$)
 }       
 $T \leftarrow$ FlowToTrajectories($G_r$)
\end{algorithm}

Next we show an online extension of the NCC algorithm.

\subsection{\edit{Online negative cycle cancellation}}
\label{sec:online_ncc}
The streaming data coming from I-24 MOTION testbed necessitates an online and memory-bounded version of Algorithm~\ref{alg:ncc}. In other words, the tracking graph $G$ is dynamic: new fragments are added and older fragments are removed from the graph constantly. A naive online extension of Algorithm~\ref{alg:ncc} is to run the same algorithm on each updated graph. However, it is inefficient because the majority of the graph remains the same and the majority of the computation on the shortest path is wasted. This opens opportunities for algorithm engineering improvements to improve performance. We show an online extension to the NCC algorithm briefly. 

% A possible improved online MCF can be achieved by modifying the tracking graph~\ref{fig:mot_graph} into an equivalent one as shown in~\cite{wang2020efficient}. Instead of connecting all the post nodes $v's$ to $t$, they will be directed back to $s$, and the original MCF problem is equivalent to finding the min-cost circulation on the modified graph $G$. The algorithm proceeds by adding fragments $\phi's$ ordered by last timestamp to the so called residual graph $G_r$ one at a time, to obtain a new graph $G_r^-$. For each iteration, we search for the least-cost negative cycle $\Gamma$ in the graph and push flow through the cycle to obtain the updated residual graph $G_r^+$. It can be proved that pushing flow through $\Gamma$, $G_r^+$, contains the min-cost circulation because the flow is feasible and no further negative cycles can be found in $G_r^+$.% (thus the optimality condition).\Yanbing{ok this part may not make sense without showing some math and definitions. what about deletion?}

The proposed online algorithm is based on the assumption that fragments are added to the graph in the order of last timestamp, which is a reasonable assumption in practice as this is the order that fragments are generated from object tracking. The online algorithm proceeds by adding each fragment $\phi_k$ to the residual graph from the previous iteration $G_{r,k-1}^+$ one at a time, to obtain a new graph $G_{r,k}^-$ (AddNode($G_{r,k-1}^-, \phi_k$)). This step adds two nodes $u_k$ and $v_k$ to the graph along with edges $(s\rightarrow u_k), (u_k\rightarrow v_k), (v_k \rightarrow s)$ and possibly additional transition edges incident to $u_k$. Then, we search for the least-cost negative cycle $\Gamma$ in $G_{r,k}^-$ (FindMinCycle($G_{r,k}^-$)) and push flow through the cycle to obtain the updated residual graph $G_{r,k}^+$. When all the fragments are processed, we output the trajectories $T$ by tracing all the cycles in the final residual graph. It can be proved that pushing flow through $\Gamma$, $G_{r,k}^+$ contains the min-cost circulation because the flow is feasible and no further negative cycles can be found in $G_r^+$. We denote the residual graph after adding $\phi_k$ at iteration $k$ to be $G_{r,k}^+$. The algorithm is shown in Algorithm~\ref{alg:ncc_online}.

\begin{algorithm}
\caption{Online NCC for MCC on a tracklet graph}
\label{alg:ncc_online}
\SetKwInput{KwData}{Input}
\SetKwInput{KwResult}{Result}
\SetAlgoLined
    \KwData{Set of fragments $\Phi=\{\phi_i\}$}
    \KwResult{Set of trajectories $T=\{\tau_i\}$}

\quad $f \leftarrow 0$

\quad $G_{r,0}^{+} \leftarrow \{s\}$

\quad $k \leftarrow 1$

\For {each $\phi_k$ \textrm{(ordered by last timestamp)}} {
    $G_{r,k}^- \leftarrow$ AddNode($G_{r,k-1}^+$ , $\phi_k$) 
    
    $\Gamma \leftarrow $ FindMinCycle($G_{r,k}^-$)
    
    $G_{r,k}^+ \leftarrow $ PushFlow($G_{r,k}^-$, $\Gamma$)
   
   $k \leftarrow k+1$
}
 $T \leftarrow$ FlowToTrajectories($G_{r,k}^+$)
\end{algorithm}

The online algorithm is proved to be correct. To summarize, we proved that the circulation in $G_{r,k}^+$ is always optimal, i.e., there is no more negative cycles in $G_{r,k}^+$ for every $k$ (optimality condition). Details on the proof can be found in~\cite{wang2023onlinemcf}.

\subsection{Trajectory rectification}

After applying fragment association, the next step is to rectify the stitched, raw trajectories with denoising, imputation and smoothing operations. Instead of ad-hoc correcting each source of the detection errors shown in Figure~\ref{fig:detection_errors}, we treat all noises and errors in a one-step approach to rectify them all at once. To simplify the problem, \edit{we consider the fixed-point positions of a trajectory, and the dimension of each vehicle is taken as the median from the bounding box trajectories. That way, the projected footprints of each vehicle in the roadway coordinate do not change shapes. }We consider a 2D vehicle motion model, with independent longitudinal ($x$) and lateral ($y$) dynamics. This allows us to decompose the problem to solving two independent 1D reconciliation problems. 
% The reconciled trajectories automatically satisfy the state consistency amongst the orders of differentiation, i.e., the finite-difference of position is speed and the finite-difference of speed is acceleration, etc.

\begin{figure}
    \centering
    \includegraphics[width = 2.5in]{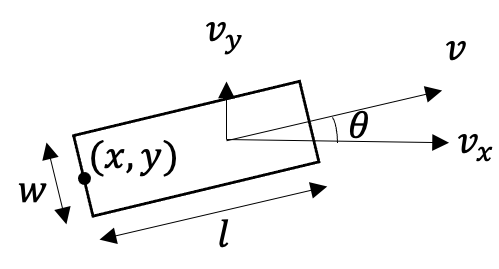}
    \caption{A simple 2D vehicle model}
    \label{fig:2d_model}
\end{figure}

We use the rear bottom-center of the 3D bounding boxes as the points of interest. Each trajectory $\tau_i$ includes the following features (features are the same for all trajectories, therefore the index $i$ is dropped for simplicity): let $\textbf{x}=[x[1], x[2],...,x[N]]^T$ be the x-positions for $N$ timesteps, and similarly let $\textbf{y}$ be the time-series of the y-position; $\textbf{v}_x = [v_x[1], v_x[2],...,v_x[N-1]]^T$ is the time-series of speed in the longitudinal direction (x-axis), and similarly $\textbf{v}_y$ is the time-series of speed in the lateral component (y-axis). The acceleration $\textbf{a}$ and jerk $\textbf{j}$ use the same 2D representation. Lastly, $l, w$ represent the vehicle length and width, respectively. The vehicle dynamics model can be seen in Figure~\ref{fig:2d_model}.

Consider the following discrete-time 3rd order 1D motion model in either longitudinal or lateral direction (the same method is applied for $x$-component and $y$-component dynamics. We demonstrate on $x$-component only):

\begin{equation}
\label{eq:discrete_dyn}
    \begin{aligned}
    x[t+1] &= x[t]+v[t]\Delta T\\
    v[t+1] &= v[t]+a[t]\Delta T\\
    a[t+1] &= a[t]+j[t]\Delta T,
    \end{aligned}
\end{equation}
with $\Delta T$ as the timestep. Notice that the finite-difference method decrements the dimension of time-series as an increase of derivative order, i.e., $\textbf{x}\in \mathbb{R}^{N}$, $\textbf{v}\in \mathbb{R}^{N-1}$, $\textbf{a}\in \mathbb{R}^{N-2}$ and $\textbf{j}\in \mathbb{R}^{N-3}$. Eq~\eqref{eq:discrete_dyn} can be written in matrix multiplication form:

\begin{equation}
    \begin{aligned}
    \textbf{v} &= D^{(1)}\textbf{x}\\
    \textbf{a} &= D^{(2)}\textbf{x}\\
    \textbf{j} &= D^{(3)}\textbf{x},
    \end{aligned}
\end{equation}
where $D^{(k)}\in\mathbb{R}^{(N-k)\times N}$ represents the $k^{th}$-order differentiation operator. For example, $k=1,2,$ and $3$ can be written as:
\[
D^{(1)}=\dfrac{1}{\Delta T}
\begin{bmatrix}
-1 & 1 & 0 & ... & 0 & 0 & 0\\
0 & -1 & 1 & ... & 0 & 0 & 0\\
\vdots & \vdots & \vdots &  & \vdots  &\vdots  & \vdots \\
 0 & 0 & 0 &... & -1 & 1 & 0\\
 0 & 0 & 0 &... & 0 & -1 & 1\\
\end{bmatrix}
\].
\[
D^{(2)}=\dfrac{1}{{\Delta T}^2}
\begin{bmatrix}
1 & -2 & 1 & 0 &... & 0 & 0 & 0\\
0 & 1 & -2 & 1 &... & 0 & 0 & 0\\
\vdots & \vdots & \vdots &  \vdots & &\vdots  & \vdots & \vdots \\
 0 & 0 & 0 & 0 &.... & 1 & -2 & 1\\
\end{bmatrix}
\]
\[
D^{(3)}=\dfrac{1}{{\Delta T}^3}
\begin{bmatrix}
-1 & 3 & -3 & 1 &... & 0 & 0 & 0 & 0\\
\vdots & \vdots & \vdots &  \vdots & &\vdots &\vdots  & \vdots & \vdots \\
 0 & 0 & 0 & 0 &.... & -1 & 3 & 3 & 1\\
\end{bmatrix}.
\]
% Assume the measurement model:
% \begin{equation}
%     \textbf{z} = H\textbf{x} + \textbf{w},
% \end{equation}
% 

% We want to find the reconstructed position $\hat{\textbf{x}}$ that is smooth in $k^{th}$-order derivatives:
% \begin{equation}
%     \label{eq:opt3}
%     \hat{\textbf{x}}=\underset{\textbf{x}}{\text{argmin}}\ 
%     \lVert\textbf{z}-H\textbf{x}\rVert^2_{2} +\delta\lVert D^{(k)} \textbf{x}\rVert_2^2 
% \end{equation}
% $\delta$ is weight. It has an analytical solution
% \begin{equation}
%     \hat{\textbf{x}} = (H^T H+\delta D^{(k) T} D^{(k)})^{-1}\textbf{z}.
% \end{equation}

Consider corrupted measurement $\textbf{z}$ which contains missing data (indicated by the observation operator $H$), noises $\textbf{w}$ and outliers $\textbf{e}$:

\begin{equation}
    \textbf{z} = H\textbf{x} + \textbf{w} + \textbf{e}, \quad \textbf{z} \in \mathbb{R}^{M}, \textbf{w} \in \mathbb{R}^{M}, \textbf{e} \in \mathbb{R}^{M}, H\in \mathbb{R}^{M\times N},
\end{equation}
and missing data exists if $M < N$ . We aim to find the reconstructed position $\hat{\textbf{x}}$ that is smooth in $k^{th}$-order derivatives. The idea is to use a combination of Ridge and Lasso regression~\cite{zou2005regularization}, to simultaneously handle noises and outliers, assuming outliers are sparse and noises have small magnitude:
% \begin{equation}
%     \label{eq:opt3_e}    (\hat{\textbf{x}},\hat{\textbf{e}})=\underset{\textbf{x},\textbf{e}}{\text{argmin}}\ 
%     \lVert\textbf{z}- H\textbf{x}-\textbf{e}\rVert^2_{2} + \sum_{k=2}^{k=K}\lambda_k \lVert D^{(k)} \textbf{x} \rVert_2^2 +\lambda_1 \lVert \textbf{e}\rVert_1.
% \end{equation}

\begin{mini!}|l|
  {\textbf{x},\textbf{e}}{\lVert\textbf{z}- H\textbf{x}-\textbf{e}\rVert^2_{2} + \sum_{k=2}^{k=K}\lambda_k \lVert D^{(k)} \textbf{x} \rVert_2^2 +\lambda_1 \lVert \textbf{e}\rVert_1}{\label{eq:opt3_e}}{\label{eq:opt3_e_obj}}
  \addConstraint{-D^{(1)}\textbf{x}\preceq 0 \label{eq:opt3_3_c1}}{}
  \addConstraint{\floor{\textbf{x}}^{(k)}\preceq D^{(k)}\textbf{x} \preceq \ceil{\textbf{x}}^{(k) }, k=2,3,...,K.\label{eq:opt3_3_c2}}{}
 \end{mini!}

The first term of the cost function~\eqref{eq:opt3_e_obj} penalizes the data-fitting error on the non-missing entries. The second term regularize the smoothness of the position vector, by penalizing the $l_2$-norm of higher-order derivatives (e.g., $k=2$ and $k=3$ correspond to accelerations and jerks in respective). The third term regularizes the sparsity of the outliers. The first constraint~\eqref{eq:opt3_3_c1} states that the speed has to be non-negative, i.e., no cars are traveling backward at any time. The second constraint~\eqref{eq:opt3_3_c2} sets the upper and lower bound for each high-order derivatives. For example, $\ceil{\textbf{x}}^{(2)}$ is the largest possible acceleration.
Note that~\eqref{eq:opt3_e_obj} can be written into a quadratic programming form by converting the $l_1$ penalization to a linear programming with linear inequality constraints~\cite{chen2001atomic}~\cite{boyd2004convex}. The problem can be solved with a convex programming solver such as \texttt{cvxopt}~\cite{vandenberghe2010cvxopt}. Note that above formulation rectifies the trajectory of each vehicle independently of another. After treatment~\eqref{eq:opt3_e}, further investigation is needed to determine if, for example, considering vehicular interactions in the optimization formulation (e.g., with additional non-collision constraint) is required.

\edit{Solving this convex program produces derivative quantities that are internally consistent, i.e., numerical differentiation of the position vector gives the speed vector, and Euler forward integration of the speed vector produces the same position vector. The same internal consistency is satisfied between other derivative orders.}

The 2D dynamics can be obtained by solving two independent optimization problems of form \eqref{eq:opt3_e} for both longitudinal and lateral dynamics. The lateral movement follows the same discrete-time dynamics as~\eqref{eq:discrete_dyn}. Let the solution for solving~\eqref{eq:opt3_e} for the longitudinal movement be $(\hat{\textbf{x}},\hat{\textbf{e}}_x)$, and the solution for the lateral movement be $(\hat{\textbf{y}},\hat{\textbf{e}}_y)$. The rectified steering angle $\hat{\boldsymbol{\theta}} = [\theta[1],...,\theta[N-1]]^T$ can be calculated in the end as

\begin{equation}
    \hat{\boldsymbol{\theta}}= \textrm{tan}^{-1}\left(\dfrac{D^{(1)}\hat{\textbf{y}}}{D^{(1)}\hat{\textbf{x}}}\right).
\end{equation}

\section{Experiments}
\label{sec:experiments}
\edit{
In this section, we present the results of four experiments designed to evaluate the performance of our proposed 2-step data reconciliation pipeline:
\begin{itemize}
    \item Experiment 1 (Exp1): In this numerical experiment, we utilize a microsimulation dataset as the ground truth. Raw tracking data is generated by intentionally degrading the ground truth to simulate upstream errors. The objective is to assess the 2-step data reconciliation pipeline's ability to reconstruct the original ground truth data from this downgraded version.
    \item  Experiment 2 (Exp2): This experiment employs a 15-minute NGSIM I-101 dataset as the ground truth. The raw tracking data is produced by manually introducing perturbations to the ground truth. The aim is to evaluate how effectively our 2-step data reconciliation pipeline can recover the ground truth in this scenario. 
    \item Experiment 3 (Exp3): Real tracking data from 16-17 cameras recorded on a section of the I-24 MOTION system serves as the input for this experiment. The data captures three distinct traffic conditions. We use manually labeled 3D vehicle bounding boxes for each scene (I24-3D~\cite{gloudemans2023interstate}) as the ground truth to validate our algorithm's performance.
    \item  Experiment 4 (Exp4): This experiment assesses the scalability of our algorithms. The input is a 4-hour raw tracking data from the full-scale I-24 MOTION system, which includes 276 cameras covering a 4.2-mile highway segment.
\end{itemize}

A summary of these experiments is provided in Table~\ref{tab:exp_description}.}

% on three microsimulation datasets, as well as on the vehicle detection and tracking outputs from two recorded videos of the I-24 MOTION system. The traffic scenarios in these datasets are sufficiently varied, and they capture representative artifacts observed in the real tracking results. We show that the postprocessing pipeline improves a variety of metrics in all the test cases.

% The organization of this section is as follows: first we describe the evaluation metrics used in all the experiments. Next we describe implementation details related to the cost function in the data association step, and the parameter values in the reconciliation step. Details for both the numerical (microsimulation) datasets as well as the real tracking datasets are described next. Finally we discuss the results and possible improvements.

% \usepackage{booktabs}
\begin{table}[h]
\centering
\resizebox{\textwidth}{!}{%
\begin{tabular}{@{}lllllll@{}}
\toprule
Experiment &
  Ground truth data &
  Raw tracking data &
  Spatial size &
  Temporal size &
  \# Cameras &
  Metrics / Goal \\ \midrule
Exp1 &
  \begin{tabular}[c]{@{}l@{}}TransModeler \\ simulation\end{tabular} &
  \begin{tabular}[c]{@{}l@{}}artificially downgraded \\ from ground truth\end{tabular} &
  2000 ft &
  15 min &
  3 &
  \begin{tabular}[c]{@{}l@{}}MOT metrics\\ Kinematic distributions\end{tabular} \\
Exp2 &
  NGSIM I-101 &
  \begin{tabular}[c]{@{}l@{}}artificially downgraded \\ from ground truth\end{tabular} &
  2200 ft &
  13.3 min &
  7 &
  \begin{tabular}[c]{@{}l@{}}MOT metrics\\ Kinematic distributions\end{tabular} \\
Exp3 &
  \begin{tabular}[c]{@{}l@{}}Manually labeled \\ I24-3D\end{tabular} &
  \begin{tabular}[c]{@{}l@{}}tracking output \\ from videos\end{tabular} &
  2000 ft &
  3 min &
  16-17 &
  \begin{tabular}[c]{@{}l@{}}MOT metrics\\ Kinematic distributions\end{tabular} \\
Exp4 &
  \begin{tabular}[c]{@{}l@{}}No available \\ ground truth\end{tabular} &
  \begin{tabular}[c]{@{}l@{}}tracking output \\ from videos\end{tabular} &
  4.5 mi &
  4 hr &
  276 &
  Scalability test \\ \bottomrule
\end{tabular}}
\caption{Summary of benchmark experiments. MOT: multi-object-tracking}
\label{tab:exp_description}
\end{table}

It is important to note that these experiments serve as an initial step towards refining the algorithms. This includes identifying optimal parameters and cost models, among others. Rather than focusing on optimizing parameters to remove all errors in a given dataset, the results provide preliminary evidence that the methods are capable of reducing errors across a range of datasets.

\subsection{Evaluation metrics}
\label{sec:metrics}
Vehicles tracking and trajectory reconciliation performance can be evaluated with standard \textit{multi-object-tracking} (MOT) metrics, specified in~\cite{bernardin2008evaluating,milan2016mot16,li2009learning,ristani2016performance}. For all the following metrics, an intersection-over-union (IOU) of vehicle footprint with ground truth of 0.3 or higher is required to be considered as a true positive.
\begin{itemize}
\setlength{\itemsep}{0pt}
    \item \textit{Precision}: number of detected objects over sum of detected and false positives (target: 1).
    \item \textit{Recall}: number of detections over number of objects (target: 1).
    \item \textit{Sw/GT}: total number of track switches per ground truth trajectory (target: 0).
    \item \textit{Fgmt/GT}: total number of fragments (switches from tracked to not tracked) per ground truth trajectory (target: 0).
    \item \textit{Multi-object-tracking accuracy} (MOTA): an aggregated measure to indicate tracking performance. It is detection errors (false negatives and false positives) and fragmentations normalized by the total number of true detections (target: 1). 
    \item \textit{Multi-object-tracking precision} (MOTP): the total error in estimated position for matched prediction and ground truth pairs over all time, averaged by the total number of matches made (target: 1).
\end{itemize}
Other statistics are computed to qualitatively assess the trajectories:
\begin{itemize}
    \item Trajectory lengths distribution
    \item Speed distribution (calculated using finite-difference)
    \item Acceleration distribution (calculated using finite-difference)
    % \item \textit{Mean absolute errors} (MAE) on position: the average absolute error of predicted position and the ground truth position for the matched instances.
\end{itemize}

\subsection{Implementation details}
In this section, we describe the computation of edge costs for min-cost flow and the choice of parameters $\lambda_i$'s for solving the trajectory rectification problem~\ref{eq:opt3_e}.

\begin{figure}
    \centering
    \includegraphics[width=2.5in]{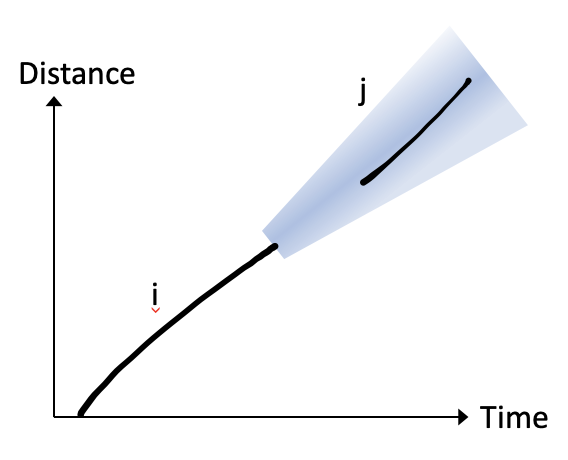}
    \caption{Probability of fragment $\phi_i$ after the last measurement of $\phi_i$ is represented as a cone. The matching cost of $\phi_i$ and $\phi_j$ is determined by the negative log likelihood of $\phi_j$ given the cone probability. }
    \label{fig:cone}
\end{figure}
The probabilities of a trajectory starts or ends with a fragment $P_{enter}$ and $P_{exit}$ is determined by the ratio of the true number of trajectories and the predicted number of trajectories. The fragment linking costs $P(\phi_i|\phi_j)$ can be modeled by (a combination of) dynamics, shape affinity and the time interval between the two fragments. We implemented a version that considers the dynamics only, as it is shown to properly represent the matching probabilities. 

The general idea is, if $\phi_j$ is indeed a continuation of $\phi_i$ but the object tracking is broken due to, e.g., object occlusion, then the projected position of $\phi_i$ at the time of $\phi_j$ has a high chance of overlapping with the actual detection of $\phi_j$. The probability of track $i$ at the time of $\phi_j$ follows a stochastic process, with the mean $\hat{p}_i$ being the projected position of track $i$ given $\phi_i$, and the variance increases with respect to $\Delta t$, the time elapsed since the end of $\phi_i$, $t^i_e$, i.e.,

\begin{equation}
p_i(t^i_e+\Delta t) = \hat{p}_i(t^i_e+\Delta t) +\eta_i(\Delta t), \textrm{ and} \ 
\eta_i(\Delta t) \sim \mathcal{N}(0, \alpha + \beta\Delta t).
\end{equation}
 
 The uncertainty $\eta_i(\Delta t)$ follows a Brownian-like process, with zero-mean and variance $\alpha+\beta \Delta t$ growing linearly with $\Delta t$ (see Figure~\ref{fig:cone}). The projected mean of position is:

\begin{equation}
\label{eq:constant_v_model}
    \hat{p}_i(t) = v_i t + \bar{p}_i,
\end{equation}
where $v_i=[v_{x,i}, v_{y,i}]^T$ and $\bar{p}_i=[x_i, y_i]^T$ can be determined using, for example, linear regression. Finally, the matching cost is the negative log likelihood of $\phi_j$ given the probabilities computed from $\phi_i$:

\begin{equation}
    \label{eq:cost}
    \Lambda(
    \phi_i, \phi_j)  = \dfrac{1}{2N_j} 
    \sum_{t_j\in[t^j_s, t^j_e]}\text{log}\left(\alpha +\beta(t_j-t^i_e)\right)+
     \dfrac{1}{2N_j} \sum_{t_j\in[t^j_s, t^j_e]}\dfrac{\left(p_j(t_j)-\hat{p}_i(t_j)\right)^2}{\alpha+\beta (t_j-t^i_e)},
\end{equation}
where $N_j$ is the number of measurements of $\phi_j$. $t^j_s$ and $t^j_e$ are the start and end timestamp of $\phi_j$, and $p_j(t_j)$ is the measurement of $\phi_j$ at $t_j$.

After fragment association, the next step is to impute missing data, correct for outliers, and smooth the trajectories in a single step by solving an optimization problem (Eq. \eqref{eq:opt3_e}). To balance the terms in the cost function, we conducted a grid search to determine the optimal parameters $\lambda_i$. We found that setting $K$ up to 3, where the $\lambda_2$ term penalizes large accelerations and $\lambda_3$ penalizes large jerks, was sufficient. Empirical assessment of a subset of trajectories led us to pre-tune the parameters to $\lambda_1 = 1.2 \times 10^{-3}$, $\lambda_2 = 1.67 \times 10^{-2}$, and $\lambda_3 = 1.67 \times 10^{-7}$. The acceleration and jerk constraints were set to $\pm 10$ ft/sec$^2$ and $\pm 10$ ft/sec$^3$, respectively.

The parameter tuning is subject to the implicit assumption of Euler forward discretization, with the current sampling frequency of 25Hz. We demonstrate the capacity the above choice of parameters to bring close the speed and acceleration to a more realistic range, but note that a detailed parameter optimization that generalizes to a range of datasets is part of our ongoing effort to operationalize the described methods.

\subsection{Exp1: Numerical experiments with microsimulation data}
\label{sec:numerical_experiments}
The proposed 2-step data reconciliation pipeline is benchmarked using a micro-simulation dataset that showcases various traffic characteristics. \edit{In the absence of ``ground truth" tracking data of sufficiently large scale to test our algorithms, numerical experiments with simulation data serve as a fundamental first step to refine algorithms and identify optimal parameters.}
The simulation dataset used in this experiment is the ground truth (dubbed as \edit{SIM-GT}) and offers a broad range of traffic scenarios, including freeflow and congested traffic with various lane-change rates. In addition, this dataset is artificially perturbed to replicate the major errors observed in real-world tracking input data, such as noise, masks, and overlapped fields of view from multiple cameras. The perturbed data is referred to as \edit{SIM-RAW}. The corresponding reconciled output after executing the 2-step postprocessing pipeline is termed as \edit{SIM-REC}. To evaluate the accuracy of the proposed method, we compare SIM-RAW and SIM-REC with the ground truth SIM-GT using standard MOT metrics along with other statistics related to traffic dynamics, as mentioned in Section~\ref{sec:metrics}.

\subsubsection{Simulation description}

\begin{figure}
    \centering
    \includegraphics[width=\textwidth]{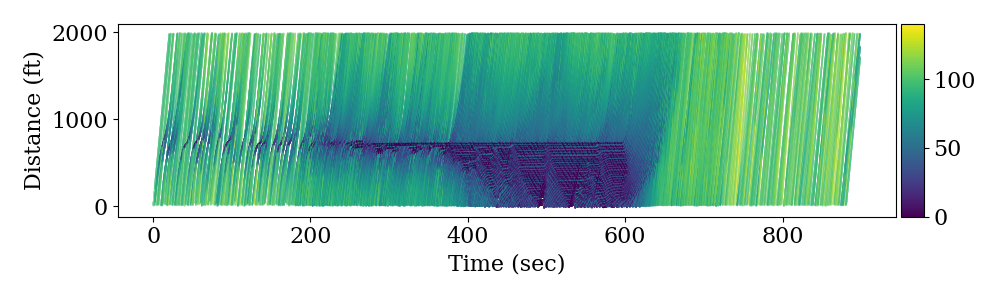}
    \caption{Ground truth simulation (SIM-GT).}
    \label{fig:SIM3}
    \centering
    \includegraphics[width=\textwidth]{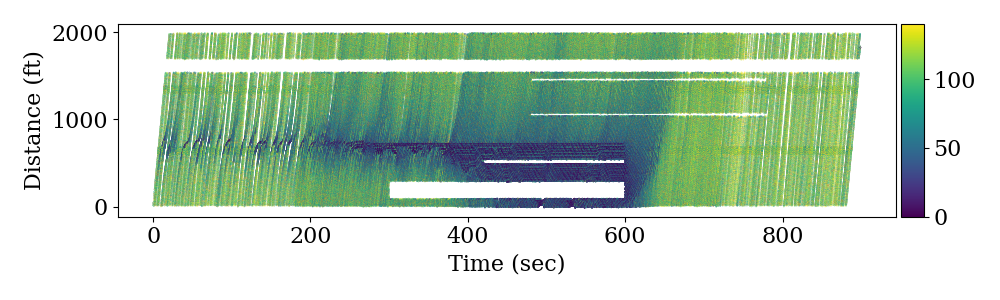}
    \caption{Manually polluted simulation data (SIM-RAW) including noises, overlapped and non-overlapped fragments.}
    \label{fig:RAW3}
    \centering
    \includegraphics[width=\textwidth]{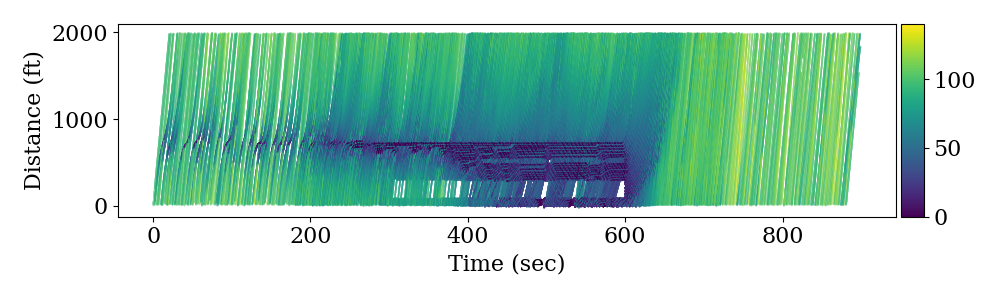}
    \caption{Reconstructed trajectory data (SIM-REC).}
    \label{fig:REC3}
\end{figure}

The simulation dataset is generated using TransModeler, a micro-simulation software that allows customization of highway traffic characteristics such as traffic density, lane blockage, car-following and lane-changing behaviors. \edit{TransModeler utilizes an extended GM model \cite{ahmed1999modeling} that closely mimics real-world car-following behaviors. The dataset encompasses a 4-lane highway segment spanning 2000 ft, with a simulation duration of 900 seconds (as depicted in Figure~\ref{fig:SIM3}). Notably, this simulation includes a lane blockage around the 800-ft mark, which persists for 400 seconds. Traffic downstream of the blockage and subsequent to its removal flows freely, while an induced congestion emerges upstream of the bottleneck.} Additional details on the simulation dataset can be found in~\ref{app:tm_data}. A summary is also provided in Table~\ref{tab:tb_sim}.

The simulation dataset has been intentionally polluted (we refer to the polluted data as SIM-RAW, see Figure~\ref{fig:RAW3}) to mimic the inaccuracies observed in the I-24 MOTION system, which arise from factors such as overpasses, misaligned cameras, and offline cameras. This introduced perturbation encompasses several elements, including spatial and temporal masks, overlapping trajectory segments, and measurement point noise.

Specifically, the SIM-RAW dataset is generated from the ground truth dataset (SIM-GT) by incorporating a mask that spans the entire simulation timeframe and covers all trajectories within the distance range of 1550-1700 ft. This mask represents an overhead bridge measuring 150 ft wide, leading to tracking disruptions for vehicles passing beneath it. Another mask is placed upstream within the distance range of 100-300 ft, where vehicles are decelerating, entering a bottleneck, and frequently changing lanes. Several additional masks, each 50 ft wide, are introduced to simulate periods of camera package loss in various locations. This results in objects changing their IDs before and after the masks.

Furthermore, we introduce a 100-ft-wide overlap in trajectories at two specific locations, occurring at distances of 600 ft and 1200 ft. These overlaps are representative of two fields of view from three cameras that intersect. The perturbation procedure also includes the introduction of random noises and outliers.

All RAW datasets have two locations (at 700 ft and 1400 ft) with overlapped regions of adjacent cameras. Vehicles change IDs when they travel across cameras, resulting in fragmentation. The fragments also overlap for about 100 ft, representing inaccuracies due to camera misalignment or homography transformation. Lastly, white noises are added to all the fragments. The resulting trajectory length, speed, and accelerations are summarized in Table~\ref{tab:tb_sim}.

% \usepackage{booktabs}
\begin{table}[h]
\centering
\begin{tabular}{@{}ll|lll@{}}
\toprule
Metrics / Statistics          &       & SIM-GT & SIM-RAW   & SIM-REC \\ \midrule
Precision $\uparrow$          &       & 1      &  0.81         &  0.98 (+21.0\%)   \\
Recall $\uparrow$             &       & 1      &  0.76         &  0.90  (+18.4\%)  \\
MOTA $\uparrow$               &       & 1      &  0.58         &  0.88 (+51.7\%)  \\
MOTP $\uparrow$               &       & 1      &  0.80         &  0.92  (+15.0\%) \\
Fgmt/GT $\downarrow$          &       & 0      &  4.66         &  0.36  (-92.3\%) \\
Sw/GT $\downarrow$            &       & 0      &  0            &  0.03 (+NA\%)  \\
No. trajectories              &       &  989  &  5597        &  1316    \\ \midrule
Trajectory lengths            & min   &  61.8 &  42.5         &  72.0    \\
(ft)                          & max   & 2001 &    802     &  2003 \\
                              & avg   & 1976 &  327.0        &  1440  \\
                              & stdev &  14.0 &  225.2   & 77.1 \\ \midrule
Speed (ft/sec)                & min   & 0.00 &  0.00   &  0.00  \\
                              & max   &  130.4  & 294.8  & 130.3 \\
                              & avg   &  57.8  &  64.1   &  60.8    \\
                              & stdev &  33.1  &  42.6     &  32.1  \\ \midrule
Acceleration                  & min   &  -6.93 &  -7608   & -6.93  \\
(ft/sec$^2$)                  & max   &  6.93 & 7446  & 6.94 \\
                              & avg   & 0.78  & 0.96     &  0.83  \\
                              & stdev &  2.80 &   1591   &  2.84   \\ \bottomrule
\end{tabular}
\caption{Simulation experiment results.}
\label{tab:tb_sim}
\end{table}
Table~\ref{tab:tb_sim} summarizes the MOT metrics for the polluted dataset. The raw data has a detection precision of 0.81, recall of 0.76, and 4.66 fragments per ground truth vehicle trajectory. We also calculate the distributions of trajectory derivative quantities as additional statistics. The polluted dataset (SIM-RAW) shows a wide spread in these metrics due to the added noises on the position. 

% \begin{figure}
%     \centering
%     \includegraphics[width=\textwidth]{figures/2023-04-15_16-02-09_transformed_beta_tm_200_gt_v4.2eb.png}
%     \caption{Time-space plot of ground truth simulation (SIM 1), with all lanes aggregated.}
%     \label{fig:SIM1}
%     \centering
%     \includegraphics[width=\textwidth]{figures/2023-04-15_16-02-35_transformed_beta_tm_200_raw_v4.2eb.png}
%     \caption{Manually polluted simulation data (RAW 1) including noises, overlapped and non-overlapped fragments.}
%     \label{fig:RAW1}
%     \centering
%     \includegraphics[width=\textwidth]{figures/2023-04-15_16-03-01_transformed_beta_tm_200_raw_v4.2__1eb.png}
%     \caption{Reconstructed trajectory data (REC 1).}
%     \label{fig:REC1}
% \end{figure}

% \begin{figure}
%     \centering
%     \includegraphics[width=\textwidth]{figures/2023-04-15_16-00-48_transformed_beta_tm_200_gt_v4.3eb.png}
%     \caption{Ground truth simulation (SIM 2).}
%     \label{fig:SIM2}
%     \centering
%     \includegraphics[width=\textwidth]{figures/2023-04-15_16-00-58_transformed_beta_tm_200_raw_v4.3eb.png}
%     \caption{Manually polluted simulation data (RAW 2) including noises, overlapped and non-overlapped fragments.}
%     \label{fig:RAW2}
%     \centering
%     \includegraphics[width=\textwidth]{figures/2023-04-15_16-01-08_transformed_beta_tm_200_raw_v4.3__7eb.png}
%     \caption{Reconstructed trajectory data (REC 2).}
%     \label{fig:REC2}
% \end{figure}

\subsubsection{Results}

The evaluation results are obtained by comparing SIM-RAW and the corresponding reconciled trajectories SIM-REC (Figure~\ref{fig:REC3}) against the same ground truth SIM-GT. The results are summarized in Tables~\ref{tab:tb_sim}. The table indicates that regardless of the traffic scenarios, the 2-step reconciliation output improves a wide range of metrics.

The proposed two-step data reconciliation approach enhances precision and recall by approximately 18-21\% when compared to the raw tracking data. This improvement is achieved by correctly associating fragments and imputing missing data caused by masks, resulting in an increase in true positives and a decrease in false negatives. False positives are reduced by merging overlapped fragments during the data-association step. Furthermore, noise is smoothed during the trajectory rectification step~\eqref{eq:opt3_e}, which reduces false positives. As a result, the aggregated score MOTA improves, while MOTP primarily improves due to the smoothing effect of the trajectory rectification step.

\begin{figure}
    \centering
    \includegraphics[width=\linewidth]{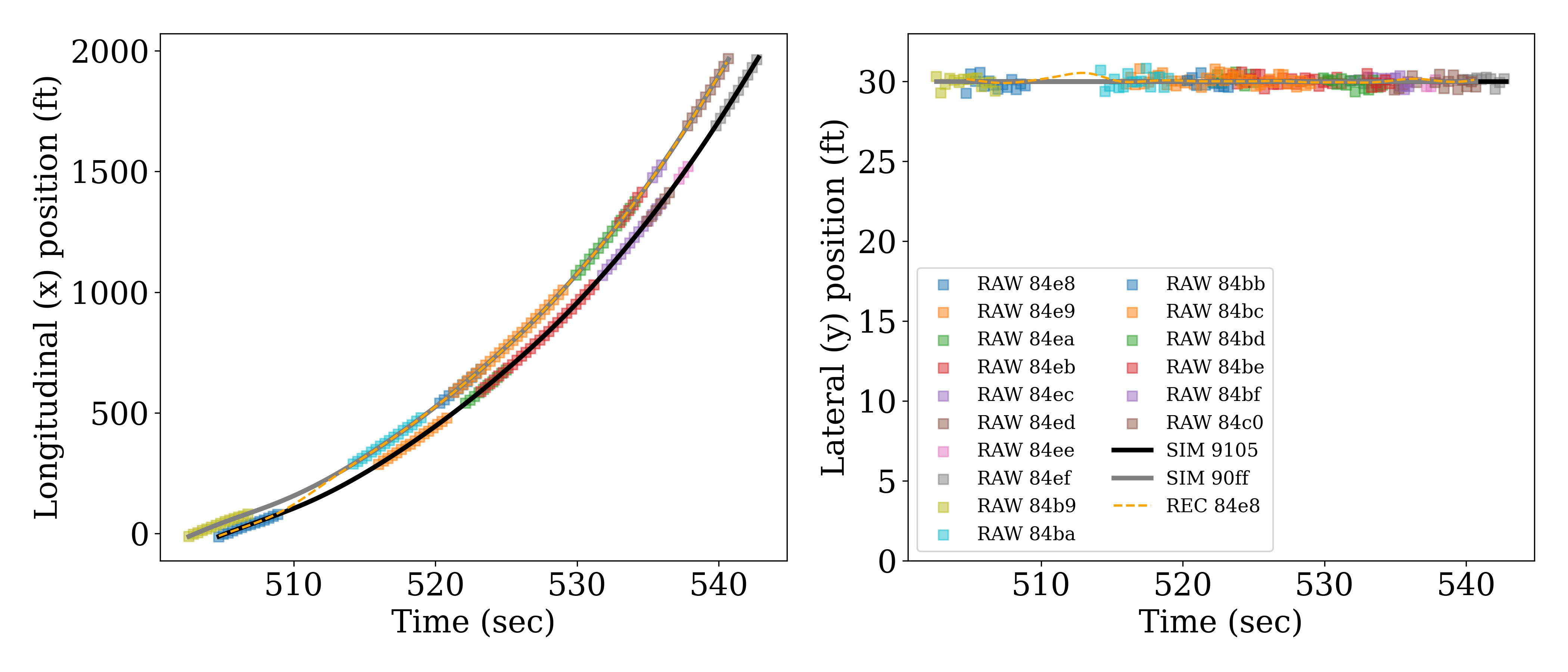}
    \caption{ID switch example 1: occluded speedup, no lane change}
    \label{fig:ids_13}
    \centering
    \includegraphics[width=\linewidth]{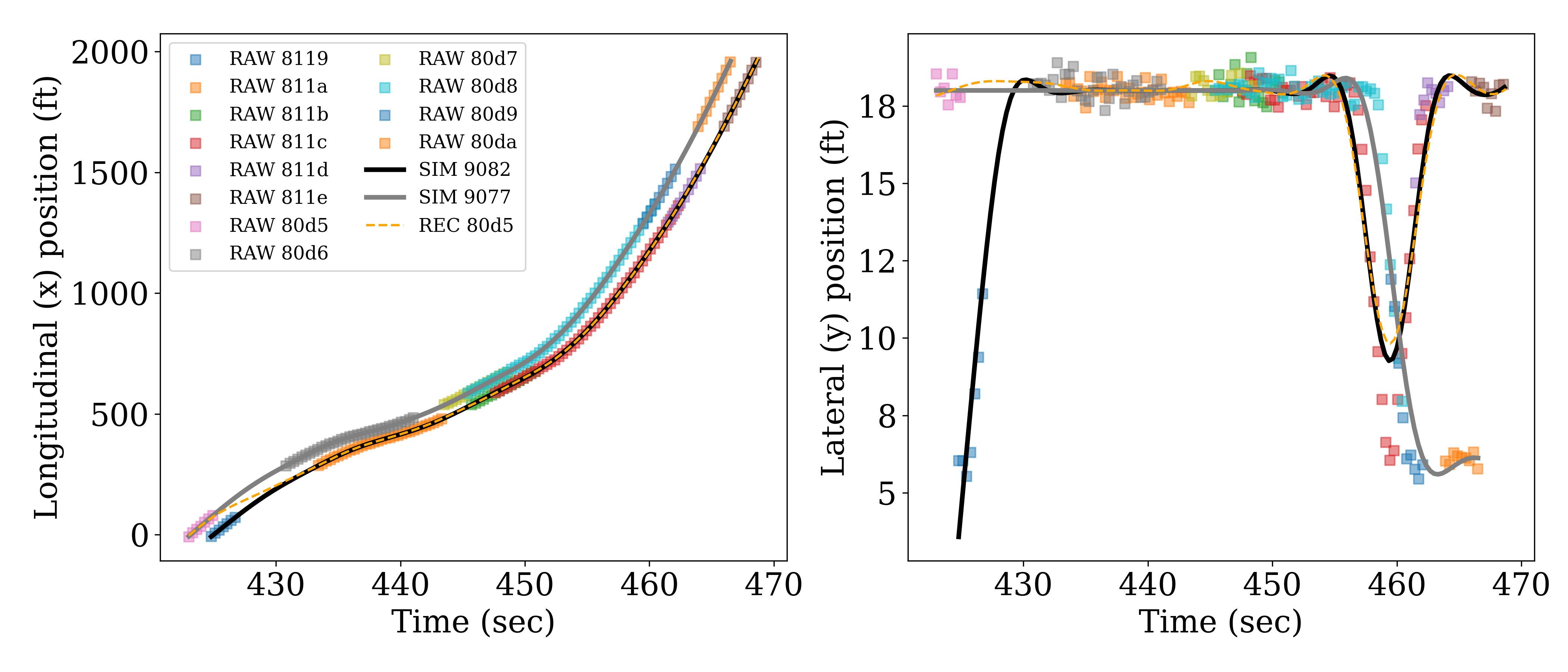}
    \caption{ID switch example 2: occluded slowdown and lane-change}
    \label{fig:ids_11}
    \centering
    \includegraphics[width=\linewidth]{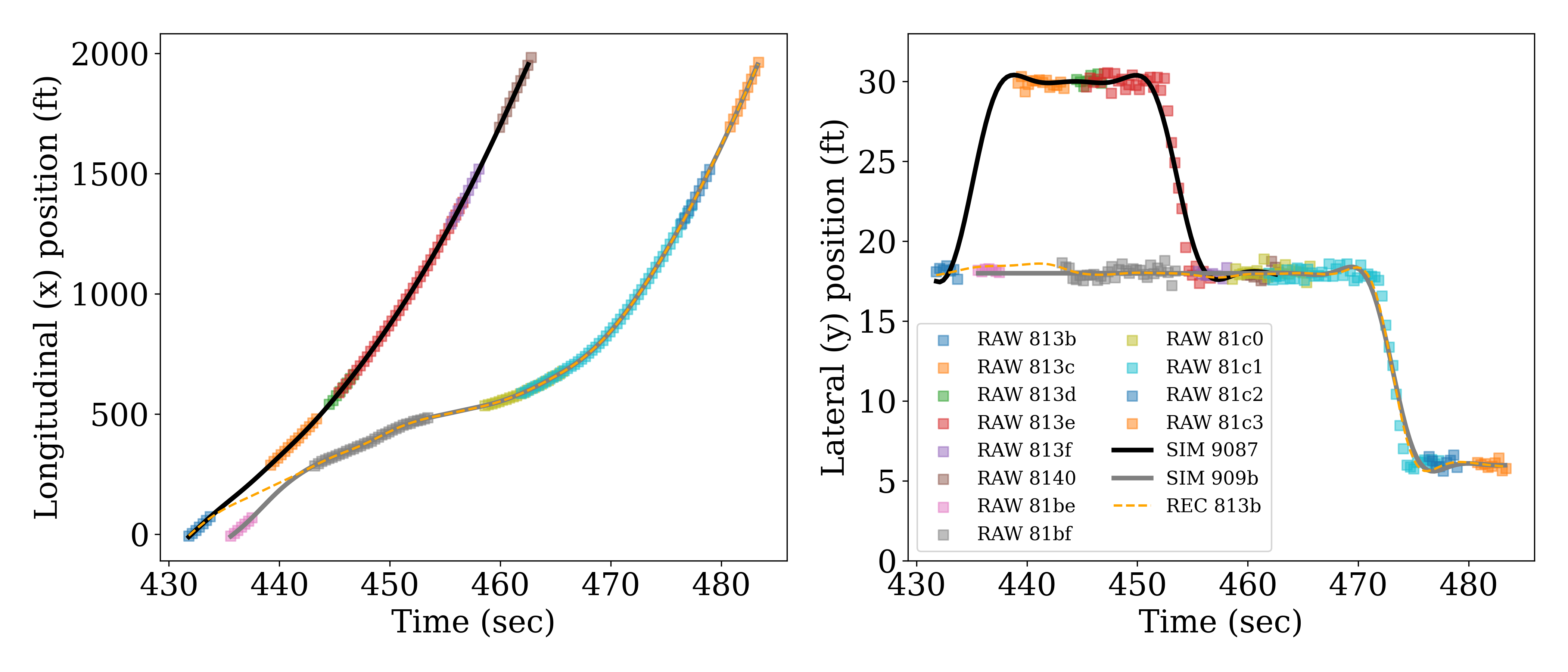}
    \caption{ID switch example 3: occluded slowdown and lane-change}
    \label{fig:ids_12}
\end{figure}

The effectiveness of the data association is particularly evident in the higher-speed regions (downstream of the bottleneck and after the lane-blockage is removed), where all fragments are correctly matched together. Overall the majority of the fragments are correctly associated (92.3\% improvement in Fgmt/GT), a few incorrectly-associated fragments lead to ID switches (i.e., when one predicted trajectory maps to multiple ground truth trajectories). In total, we counted 30 ID switches in SIM-REC and identified two reasons for their occurrence: (1) a change in speed occluded by the mask, and/or (2) an abrupt lane-change maneuver occluded by the mask. Several illustrative examples of ID switches are shown in Figure~\ref{fig:ids_13}-\ref{fig:ids_12}. 

Each plot displays a single predicted trajectory from SIM-REC that mapped to two ground truth trajectories, along with the corresponding raw fragments from SIM-RAW. The left sides show time vs. longitudinal (x) positions of all trajectories, and the right sides are time vs. lateral (y) positions. The matched tracking trajectory (SIM-REC) is shown in an orange dashed line, while the two ground truth trajectories (in SIM-GT) are depicted in black and grey solid lines. All the fragments (in SIM-RAW) corresponding to those two ground truth trajectories are displayed in colored points, with each fragment represented by a distinct color. Figure~\ref{fig:ids_13} is an example where an ID switch occurs due to an unobserved speedup at the masked region (longitudinal position 100-300 ft). Before entering the mask, both vehicles 84e8 and 84b9 traveled on the same lane at approximately 25.5 ft/s with 84e8 following closely behind 84b9. Tracking for both vehicles was discontinued for 8 seconds due to the mask, and when they reappeared at position 300 ft, the speed for both vehicles had increased to 42.8 ft/s. Our current cost function (Eq~\eqref{eq:cost}) assumes linear longitudinal and lateral dynamics, and thus favors matching two fragments that appear to be on a straight line in the time-space dimension. Consequently, fragment 84e8 has a lower cost to be matched with fragment 84ba. Matching becomes even more challenging when both longitudinal and lateral dynamics change during the occlusion, as shown in Figure~\ref{fig:ids_11} and \ref{fig:ids_12}. In such cases, the lateral dynamics tend to dominate the association cost when fragments are ``close" together in time vs. x dimension, causing the data association to fail because it prefers to match fragments that are in the same lane.

These examples highlight the difficulty of accurate motion modeling, particularly during non-trivial maneuvers such as speed changes and lane changes in long occlusions. This difficulty motivates the search for other representations of ``similarity" in matching cost design that do not solely rely on motion information. One potential solution is to use appearance embedding from the upstream object detector. Another option is to train a classifier-like similarity measure model that learns the correct matching function from labeled data. For future work, we could explore a general data-driven approach outlined in~\cite{mathisen2020learning}.

Moreover, we compute the statistics on the dynamical information of the trajectories as some unsupervised measures of data quality when ground truth labels are not present. Specifically, we analyze the statistical characteristics of trajectory dynamics. Ideally, the length distribution of trajectories should cover the entire road segment (excluding ramps or trajectories at the time boundaries). The speed distribution should reflect the traffic dynamics, including free flow, stop-and-go, and congestion, and the acceleration distribution should be within a reasonable range. We calculate speed and acceleration by taking the finite difference of positions and speed, respectively. A comparison of the traffic dynamics distributions for ground truth, polluted, and postprocessed data is presented in Table~\ref{tab:tb_sim}. Generally, SIM-REC generates longer trajectories due to the fragment association step. Notably, the regularization terms in~\eqref{eq:opt3_e} of data reconciliation contribute to a significantly narrower (or more consistent) and feasible acceleration distribution. Table~\ref{tab:tb_sim} suggests that the statistics for speed and acceleration of the reconstructed data SIM-REC are much closer to the ground truth than those of the polluted datasets.

% \input{tb_unsup}

% \begin{figure}
%     \centering
%     \includegraphics[width=\linewidth]{figures/dist_transmodeler_tm_200_raw_v4.2_V_reconciled_tm_200_raw_v4.2__1.png}
%     \caption{Distributions of trajectory lengths (left), speed (middle) and acceleration (right) for RAW 1 (blue) and REC 1 (orange).}
%     \label{fig:dist_1}
%     \centering
%     \includegraphics[width=\linewidth]{figures/dist_transmodeler_tm_200_raw_v4.3_V_reconciled_tm_200_raw_v4.3__7.png}
%     \caption{Distributions of trajectory lengths (left), speed (middle) and acceleration (right) for RAW 2 (blue) and REC 2 (orange).}
%     \label{fig:dist_2}
%     \centering
%     \includegraphics[width=\linewidth]{figures/dist_transmodeler_tm_900_raw_v4.1_V_reconciled_tm_900_raw_v4.1__0.png}
%     \caption{Distributions of trajectory lengths (left), speed (middle) and acceleration (right) for RAW 3 (blue) and REC 3 (orange).}
%     \label{fig:dist_3}
% \end{figure}

\subsection{\edit{Exp2: Experiments on the NGSIM I-101 data}}
\begin{figure}
    \centering
    \includegraphics[width=\textwidth]{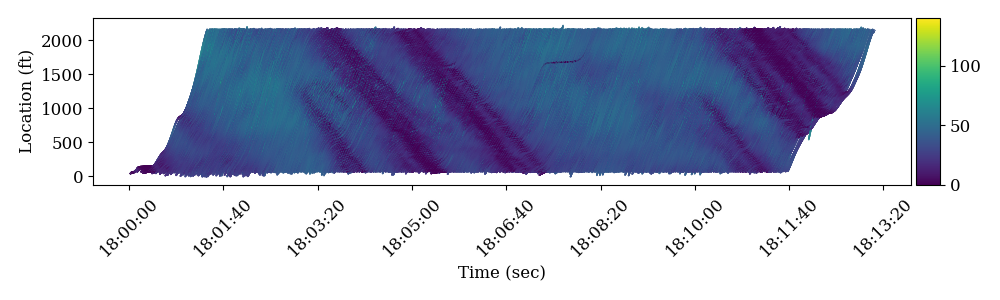}
    \caption{Ground truth NGSIM-GT.}
    \label{fig:ngsim_gt}
    \centering
    \includegraphics[width=\textwidth]{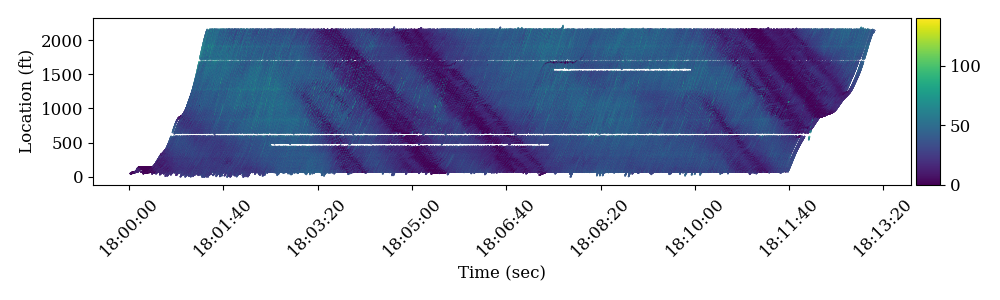}
    \caption{Manually polluted NGSIM-RAW including noises and masks.}
    \label{fig:ngsim_raw}
    \centering
    \includegraphics[width=\textwidth]{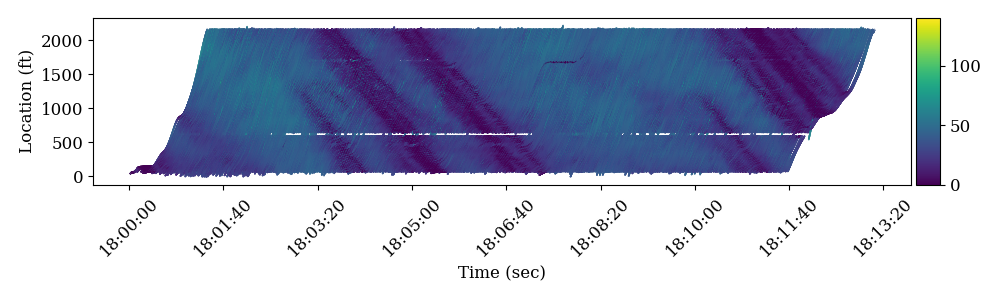}
    \caption{Reconstructed trajectory data NGSIM-REC.}
    \label{fig:ngsim_rec}
\end{figure}

Similar to Exp1, in this experiment we run our algorithms on the manually polluted NGSIM data, and benchmarked it against the ground truth NGSIM data~\cite{NGSIM2006}. The manual perturbation includes masks of widths between 30 and 50 ft, noises and outliers on position measurements. The metrics for ground truth, perturbed and reconstructed data are summarized in Table~\ref{tab:ngsim_res}, and the time-space diagrams are shown in Figure~\ref{fig:ngsim_gt}-\ref{fig:ngsim_rec}.

% \usepackage{booktabs}
\begin{table}[h]
\centering
\begin{tabular}{@{}ll|lll@{}}
\toprule
Metrics / Statistics          &       & NGSIM-GT & NGSIM-RAW   & NGSIM-REC \\ \midrule
Precision $\uparrow$          &       & 1      &  0.48         &  0.69 (+43.8\%)   \\
Recall $\uparrow$             &       & 1      &  0.46         &  0.67  (+45.7\%)  \\
MOTA $\uparrow$               &       & 1      &  0.81         &  0.86 (+6.17\%)  \\
MOTP $\uparrow$               &       & 1      &  0.84         &  0.90  (+7.14\%) \\
Fgmt/GT $\downarrow$ &       & 0      &  8.93         &  1.52  (-83.0\%) \\
Sw/GT $\downarrow$  &       & 0      &  2.69         &  0.96 (-64.3\%)  \\
No. trajectories              &       &  1533  &  13685        &  2335    \\ \midrule
Trajectory lengths            & min   &  877.7 &  3.90         &  17.7    \\
(ft)                          & max   & 2173.4 &  1049.8       &  2172.9  \\
                              & avg   & 2037.6 &  517.9        &  1030.6  \\
                              & stdev &  176.4 &  322.8        &  605.6   \\ \midrule
Speed (ft/sec)                & min   & 0.029  &  7.11         &  0.00    \\
                              & max   & 81.6   &  158.9        &  81.5    \\
                              & avg   &  29.8  &  29.94        &  29.9    \\
                              & stdev & 13.9   &  15.8         &  14.0    \\ \midrule
Acceleration                  & min   &  -6.82 &  -1784.4      &  -6.91   \\
(ft/sec$^2$)                  & max   & 6.80   &  2060.8       &  6.94    \\
                              & avg   & 0.113  &  0.132        &  0.11    \\
                              & stdev &  1.82  &  265.8        &  2.06    \\ \bottomrule
\end{tabular}
\caption{NGSIM experiment results.}
\label{tab:ngsim_res}
\end{table}

We observe that compared to NGSIM-RAW, NGSIM-REC improves on all the metrics. Particularly, 83\% of the fragments are reduced. We notice a similar result as compared to the previous experiment: the accuracy of data association is particularly high when the traffic is at a higher speed. Whereas when the speed variation is occluded by the masks, it is much difficult to match fragments that are interrupted by the masks. Additionally, we observe the speed and acceleration distributions of the reconstructed data are closer to the ground truth.

\subsection{Exp3: Experiments on the I-24 MOTION validation system}
Next, we demonstrate the application of the proposed pipeline on the I-24 MOTION validation system, which was built in 2020 and functioned as a prototype for the design selections in the full system. The validation system consists of three poles that host 18 cameras to seamlessly cover 2000-ft highway segment. The detailed description of the design can be found in~\cite{gloudemans202324}.

% The I-24 MOTION testbed~\cite{gloudemans202324} uses 276 densely instrumented high-resolution cameras to seamlessly cover a stretch of 4.2 miles, 8-lane highway on Interstate-24 near Nashville, Tennessee. 

\subsubsection{Dataset description}
\edit{
All evaluations have been conducted using three datasets sourced from the I-24 MOTION validation system (I24-3D)~\cite{gloudemans2023interstate}, each describes a distinct traffic scenario. The ground truth datasets comprise a total of 877,000 manually-labeled 3D bounding boxes of vehicles, derived from 57 minutes of video data collected across 16-17 cameras. Notably, this dataset stands as the most extensive ``ground truth" dataset directly obtained from the I-24 MOTION system's cameras. It captures real-world traffic scenes and includes raw tracking results produced by the system's current tracker. This data presents real-world challenges originating from upstream sources, including issues such as fragmentations due to occlusions, object handoffs between multiple cameras, projection errors when transitioning from camera image space to roadway coordinates, false positives arising from tracking and detection errors, and more.

The three distinct scenes in the I24-3D include:
\begin{itemize}
    \item A 60-second free-flow traffic scenario.
    \item A 51-second slow traffic in snowy conditions.
    \item A 50-second scene of heavily congested traffic with stop-and-go waves.
\end{itemize}

The raw tracking data was obtained using a crop-based fast tracking algorithm described in~\cite{gloudemans2021vehicle}. The raw tracking results obtained from the three video recordings are denoted as RAW-i, RAW-ii, and RAW-iii. The corresponding ground truth datasets are labeled as GT-i, GT-ii, and GT-iii, respectively. The output generated by our 2-step reconciliation pipeline is designated as REC-i, REC-ii, and REC-iii.}

\begin{table*}[h]
\resizebox{\textwidth}{!}{%
\centering
\begin{tabular}{@{}ll|lll|lll|lll@{}}
\toprule
Metrics &  & GT-i & {RAW-i} & REC-i & GT-ii & {RAW-ii} & REC-ii & GT-iii & RAW-iii & REC-iii\\ \midrule
Precision $\uparrow$          &       & 1      & 0.71   & 0.90   & 1    & 0.87     & 0.88  & 1  &0.76  &0.76 \\
Recall $\uparrow$             &       & 1      & 0.56   & 0.83   & 1    & 0.55     & 0.79  & 1  &0.46  & 0.67\\
MOTA $\uparrow$               &       & 1      & 0.32   & 0.74   & 1    & 0.48     & 0.68  & 1  &0.31  & 0.46\\
MOTP $\uparrow$               &       & 1      & 0.63   & 0.73   & 1    & 0.72     & 0.75  & 1  &0.67  & 0.68\\
Fgmt/GT $\downarrow$          &       & 0      & 5.22   & 0.60   & 0    & 5.38     & 1.93  & 0  &4.92  & 1.11\\
Sw/GT $\downarrow$            &       & 0      & 1.43   & 0.04   & 0    & 2.98     & 0.53  & 0  &3.01  & 0.52\\
No. trajs                     &       & 314    & 789    & 321    & 100  & 411      & 150 & 253  &1250  & 282\\
\midrule
Trajectory            & min   & 25.6  & 24.0    & 17.6   &   36.9    & 5.1     & 5.1     & 5.0     & 0.01    & 0.45\\
lengths (ft)          & max   & 2270.5 & 2096.5  & 2094.7  & 2183.8 & 568.9   & 2181.2  & 2278.1  & 2066.9  & 2289.0\\
                      & avg   & 1635.4 & 507.8   & 1455.1  & 1013.5 & 147.9   & 588.2   & 1042.0  & 157.8   & 721.2\\
                      & stdev & 639.0  & 419.4   & 711.6   & 587.9  & 126.8   & 512.2   & 621.5   & 231.8   & 649.1\\
\midrule
Speed                 & min   & 72.3   & 0.97     & 72.7    & 16.7   & 0.01   &  8.7    &  0.00  &  0.00  & 0.00\\
(ft/sec)              & max   & 147.7  & 437.8   & 142.9   & 62.0  &  276.4  &  66.36   &  136.3 &  235.2 & 141.2 \\
                      & avg   & 106.0  & 106.4   & 105.9   & 39.4   & 39.1   &  9.3     &  37.6  & 40.6   & 37.7\\
                      & stdev & 10.9   & 13.8    & 10.5    & 7.24   & 11.6   &  7.40    &  33.9  & 35.5   & 33.4\\
\midrule
Acceleration          & min   & -5.6   &  -54155      & -6.9       &  -6.58  & -20965    & -6.94  &  -6.90  & -19233 &  -6.94 \\
(ft/sec$^2$)          & max   & 4.95   & 55701.0   & 6.7      &  6.86  & 25017.1   & 6.94   & 6.90    & 54174  & 6.93\\
                      & avg   & -0.005   & -9.76     & 0.12     &  -0.04  & 39.2      & -0.04  & -0.157  & 9.1    &  -0.159  \\
                      & stdev & 1.29  & 1891.3    & 1.46     &  1.01  & 675.8     & 1.48   & 2.16    & 1002.3 &  2.21\\ \bottomrule
\end{tabular}}
\caption{Exp2: Evaluation results using 3 manually-labeled ground truth datasets~\cite{gloudemans2023interstate}. }
\label{tab:iccv_res_combined}
\end{table*}
% \begin{figure}
%     \centering
%     \includegraphics[width=\linewidth]{figures/dist_trajectories_ICCV_raw1_V_reconciled_ICCV_raw1__8.png}
%     \caption{Distributions of trajectory lengths (left), speed (middle) and acceleration (right) for RAW-i (blue) and REC-i (orange).}
%     \label{fig:dist_i}
%     \centering
%     \includegraphics[width=\linewidth]{figures/dist_trajectories_ICCV_raw2_V_reconciled_ICCV_raw2__0.png}
%     \caption{Distributions of trajectory lengths (left), speed (middle) and acceleration (right) for RAW-ii (blue) and REC-ii (orange).}
%     \label{fig:dist_ii}
% \end{figure}

\subsubsection{Results}
The performance of the pipeline on the three tracking datasets is presented in Table~\ref{tab:iccv_res_combined}. The results show that the proposed data reconciliation pipeline improves on all the MOT metrics for all three scenes. Noticeably the pipeline is able to match the 60-77\% of fragments. The data association step enhances precision and recall by accurately connecting fragments and filling in the gaps between the detections. 
% Furthermore, the smoothing step leads to a more realistic range of speed and acceleration, as compared to the raw tracking data which contains noises. Figure~\ref{fig:dist_i} and \ref{fig:dist_ii} provide a comparison of trajectory lengths, speed, and acceleration distributions before and after postprocessing.

\begin{figure}[]
    \centering
    \includegraphics[width=\linewidth]{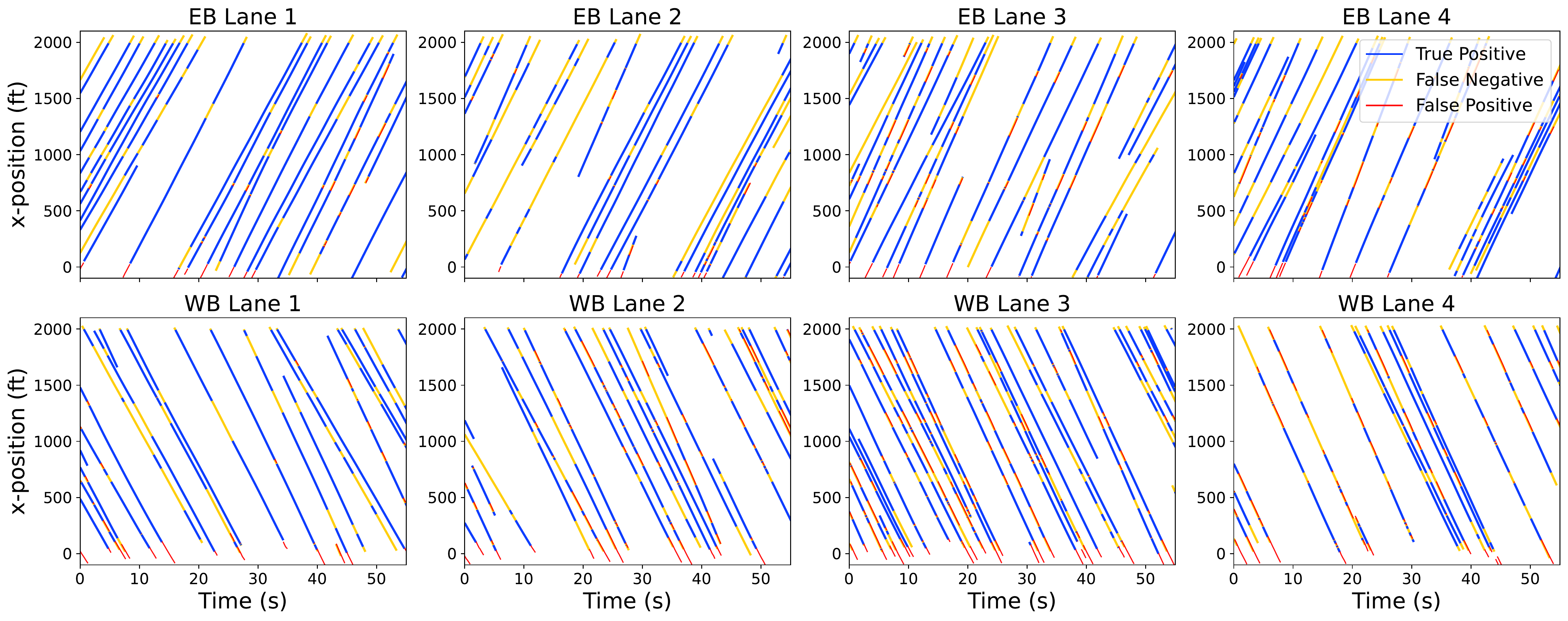}
    \caption{Lane-specific time-space diagrams for RAW-i (Lane 4 is rightmost lane in direction of travel). False negatives (yellow), false positives (red) and true positives (blue) are shown. False negatives indicate missing data due to disrupted tracking, and false positives indicate that an object was tracked below the IOU threshold. }
    \label{fig:ts1_eval}
    \includegraphics[width=\linewidth]{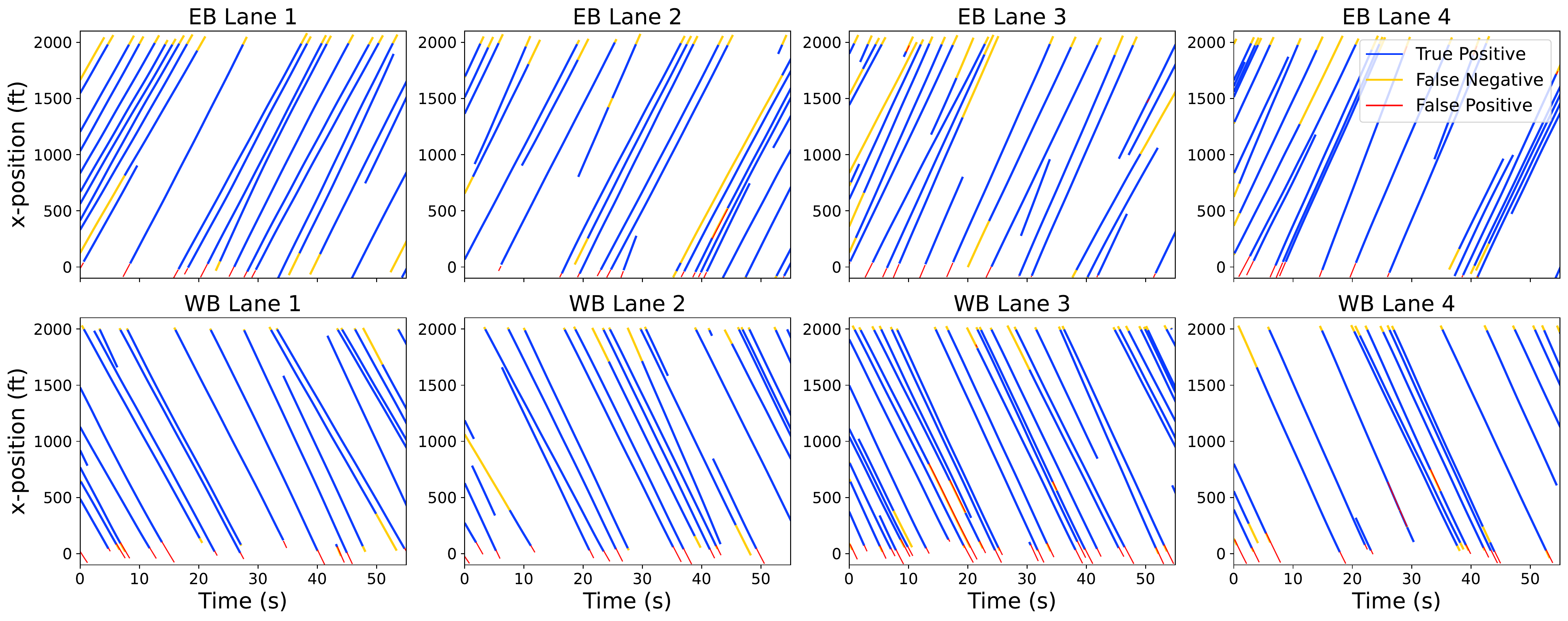}
    \caption{Lane-specific time-space diagrams for REC-i. False negatives are significantly reduced due to data association and imputation.}  
    \label{fig:ts1_eval_rec}
\end{figure}

\begin{figure}[]
    \centering
    \includegraphics[width=\linewidth]{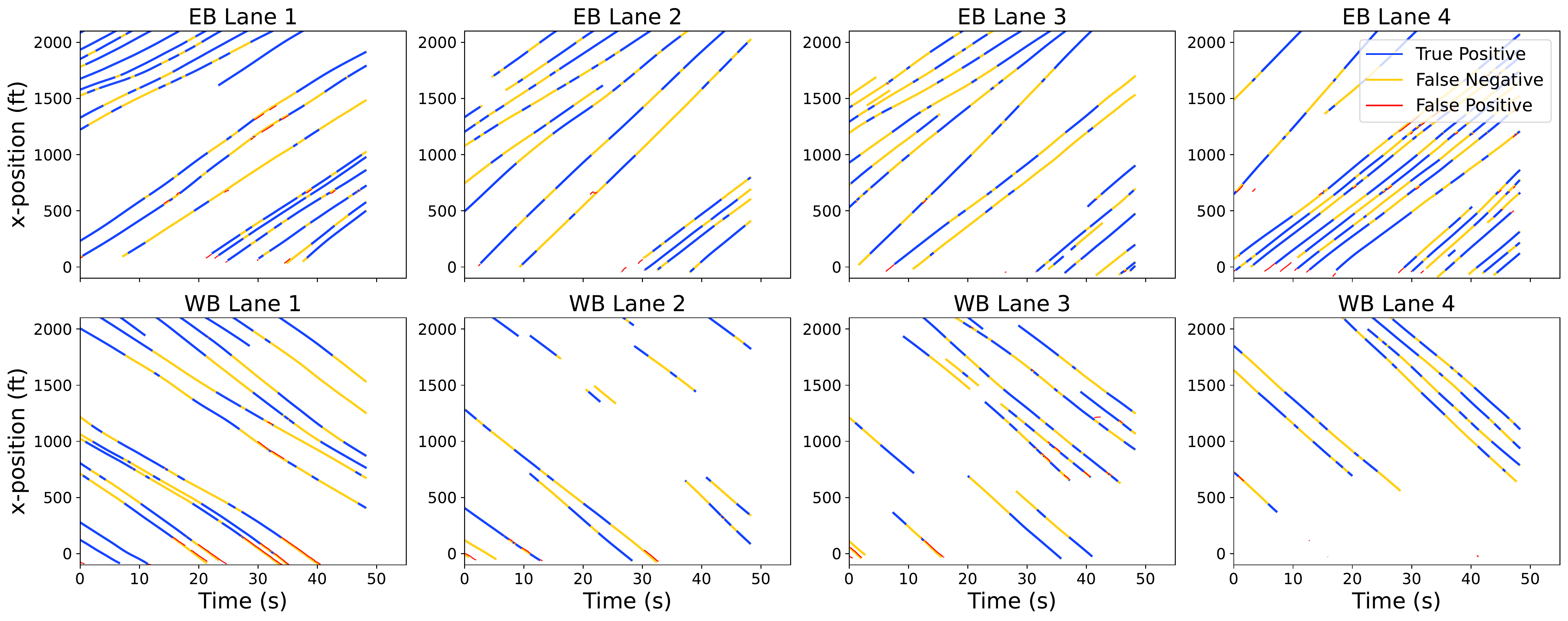}
    \caption{Lane-specific time-space diagrams for RAW-ii (snow, low speed traffic). False negatives dominate the errors.}
    \label{fig:ts2_eval}
    \includegraphics[width=\linewidth]{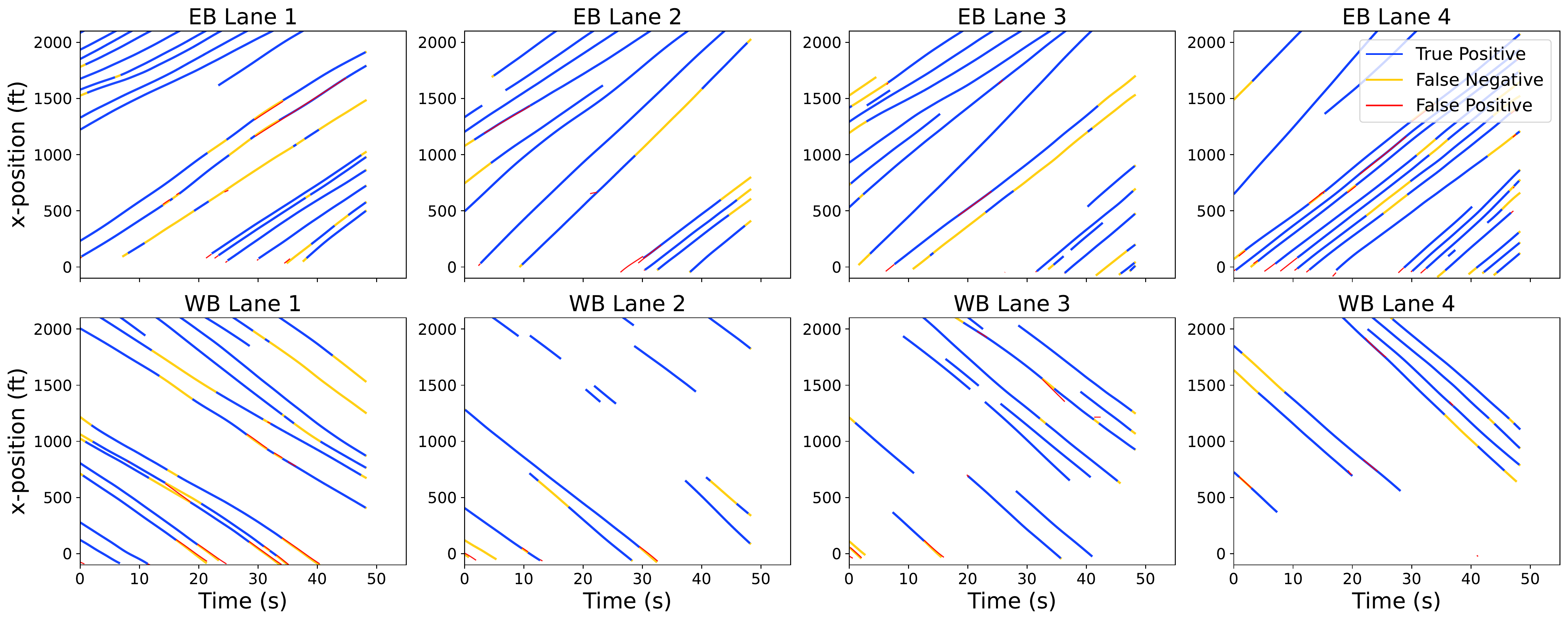}
    \caption{Lane-specific time-space diagrams for REC-ii. False negatives are reduced.}
    \label{fig:ts2_eval_rec}
\end{figure}

\begin{figure}[]
    \centering
    \includegraphics[width=\linewidth]{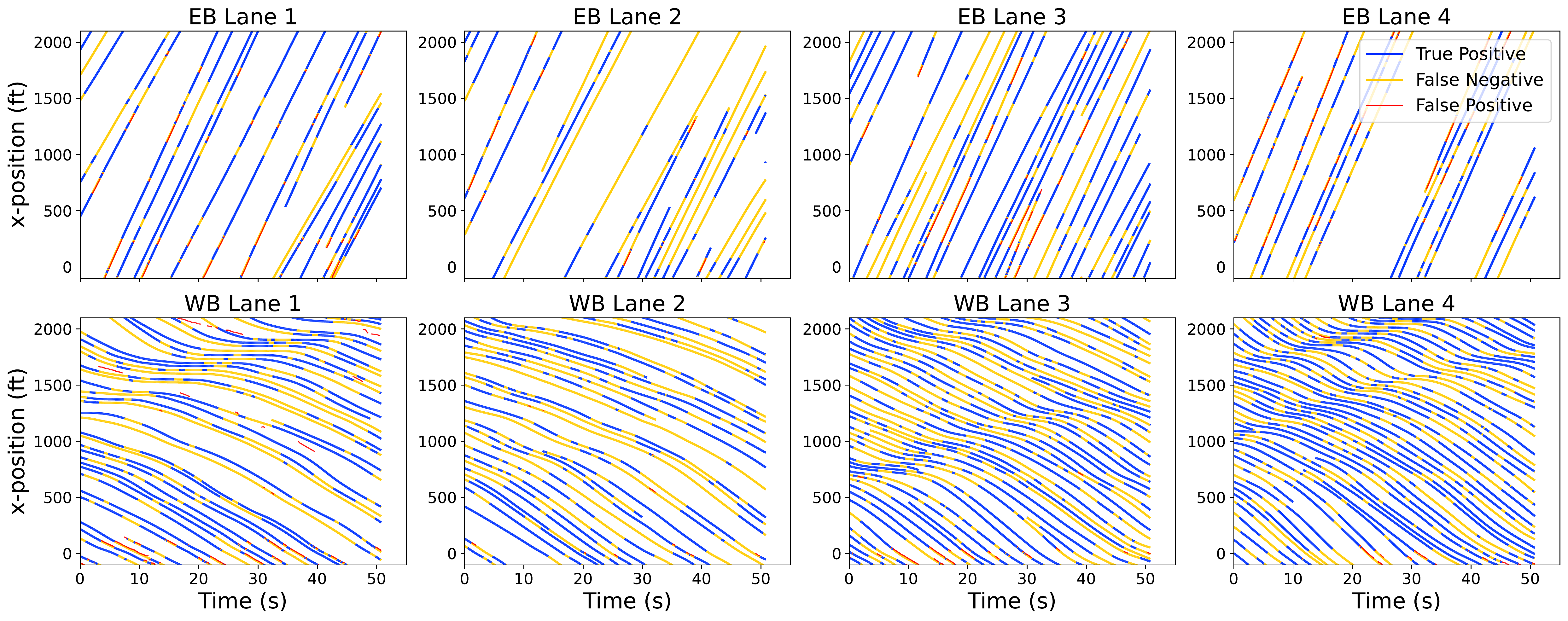}
    \caption{Lane-specific time-space diagrams for RAW-iii (heavy congestion in WB). WB tracking is extremely fragmented.}
    \label{fig:ts3_eval}
    \includegraphics[width=\linewidth]{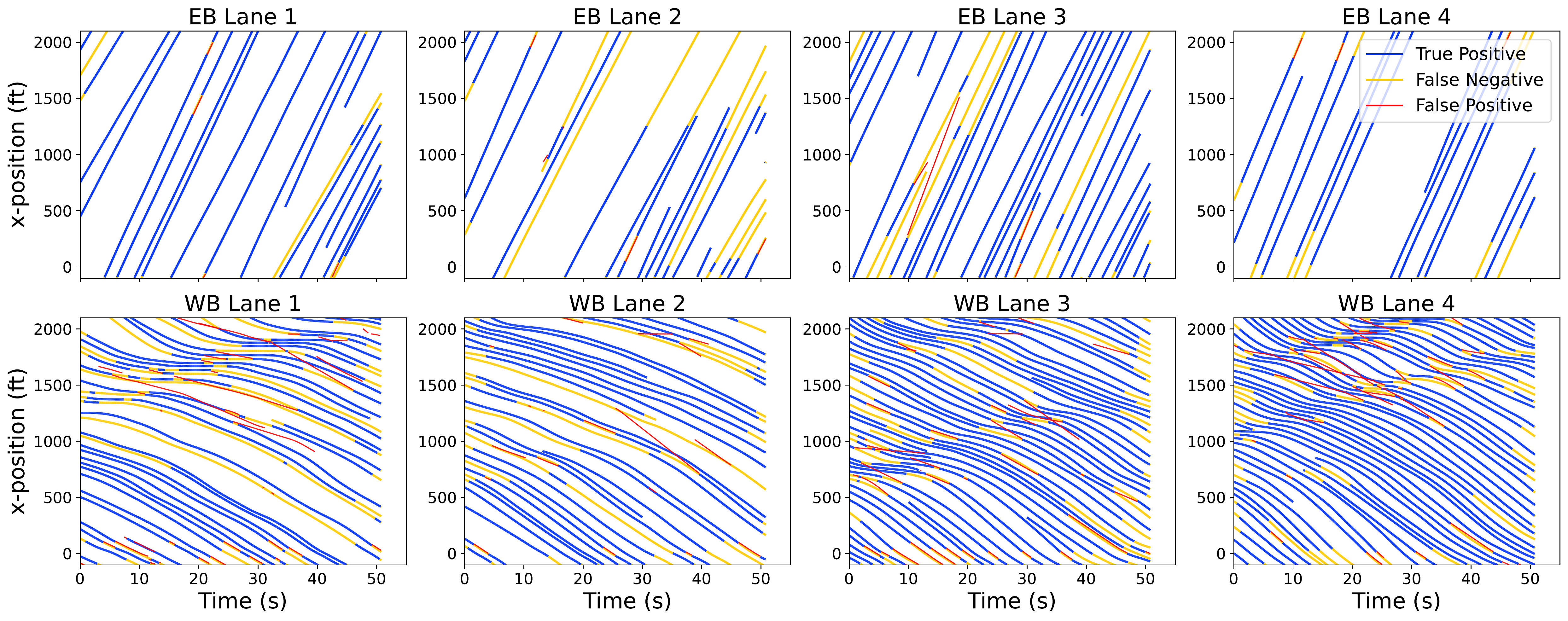}
    \caption{Lane-specific time-space diagrams for REC-iii. Fragmentation is addressed in most cases. However, ID switching can occur when the speed varies, due to the constant-velocity assumption in the current matching cost formulation. }
    \label{fig:ts3_eval_rec}
\end{figure}

\edit{To view the performance in more details, Figures~\ref{fig:ts1_eval}-\ref{fig:ts3_eval_rec} provide the time-space diagram for each lane colored by true positives, false positives, false negatives. The results show comparison between the raw tracking (RAWs) and the ground truth (GTs), as well as the postprocessed (RECs) and GTs. Specifically, we show that in all three scenes, the false negatives are significantly reduced due to the data association and the data imputation in between the associated fragments. The false negatives at the boundaries remain after postprocessing because there is currently no mechanism to extrapolate trajectories to the boundaries of the time window. A few false positives are corrected automatically during the data association stage, and others are identified as outliers and are removed at the trajectory rectification stage. Most significantly, we show that most of the the ``choppy" fragments in  RAW-iii westbound (Figure~\ref{fig:ts3_eval_rec}) are correctly associated together, leading to more complete trajectories. The fragmentation rate reduced from 4.92 to 1.11 Fgmt/GT. However, postprocessing is likely to reintroduce false positives if two fragments are matched incorrectly (causing ID switching). This error tend to occur when a vehicle undergoes speed variations and the tracking is fragmented, as the case in REC-iii westbound. The overall tracking and postprocessing accuracy is the highest in REC-i (freeflow), as the vehicle dynamics are the simplest.  Overall, the benchmark experiments demonstrate a promising result that the postprocessed trajectories resolved the majority of the fragmentation issues, and significantly reduced false negatives. }

\begin{figure}
    \centering
    \includegraphics[width=0.9\linewidth]{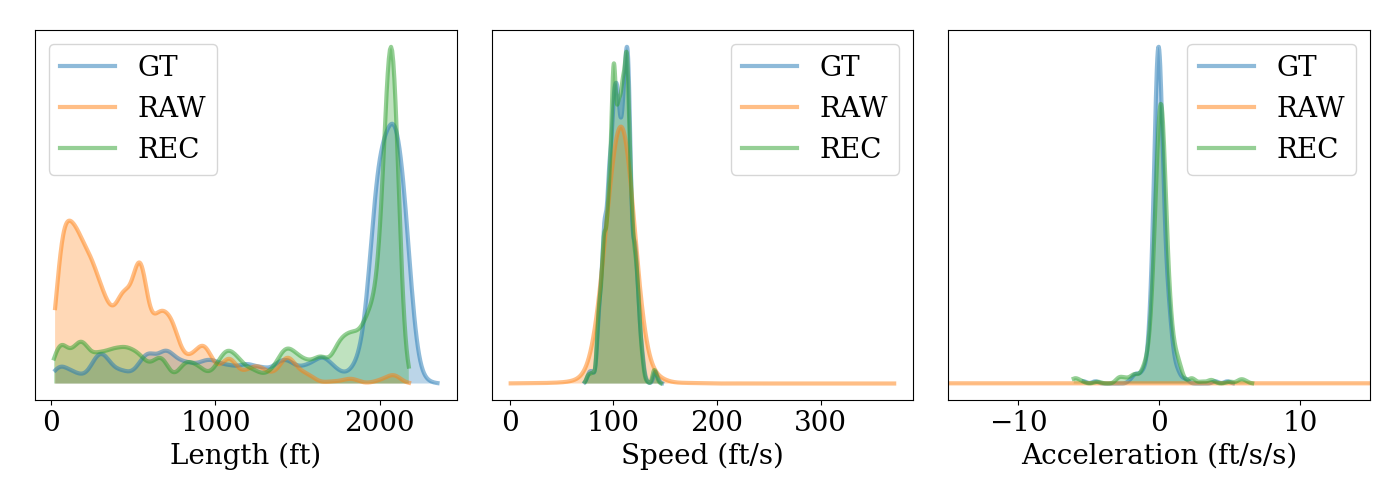}
    \caption{Distributions of trajectory lengths (left), speed (middle) and acceleration (right) for GT (blue), RAW (orange) and REC (green) for scene i (free flow).}
    \label{fig:dist_trajectories_ICCV_gt1}
    \includegraphics[width=0.9\linewidth]{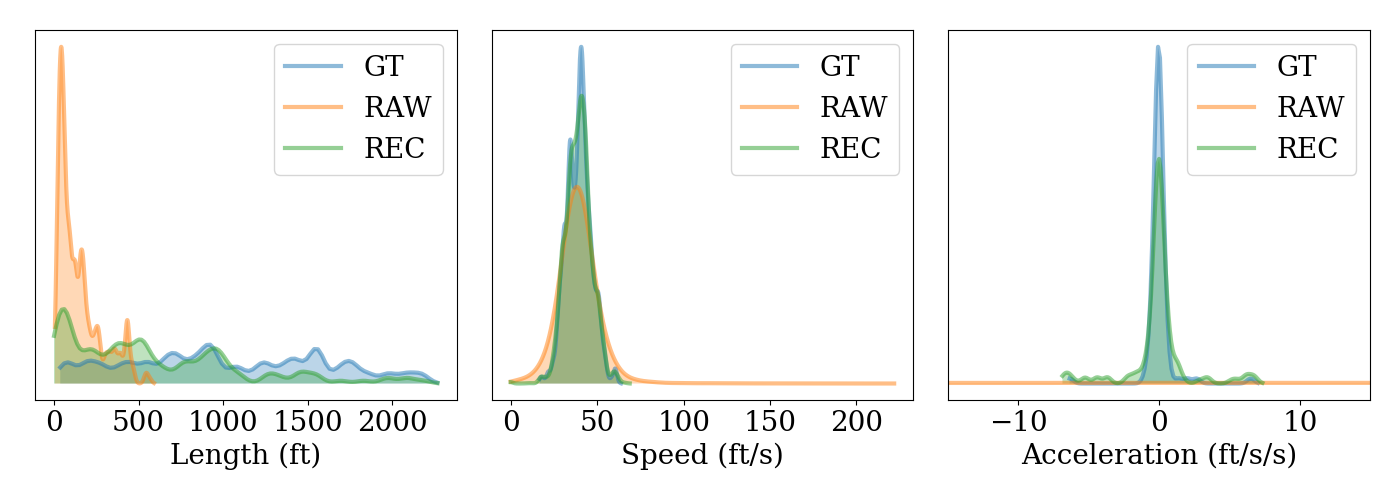}
    \caption{Distributions of trajectory lengths (left), speed (middle) and acceleration (right) for GT (blue), RAW (orange) and REC (green) for scene ii (snow weather and slow traffic).}
    \label{fig:dist_trajectories_ICCV_gt2}
    \includegraphics[width=0.9\linewidth]{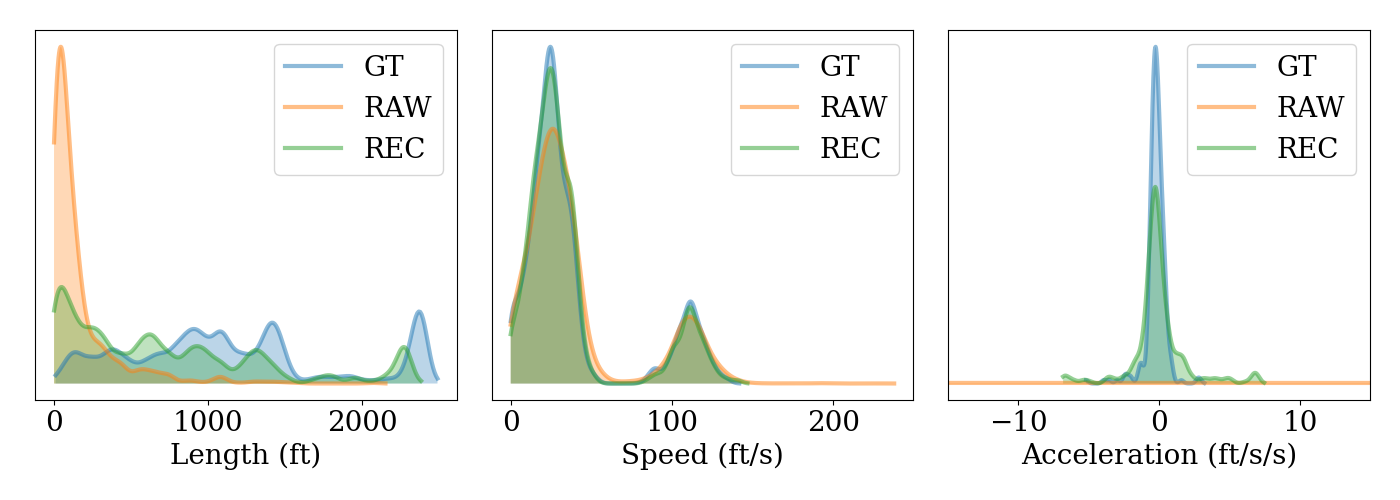}
    \caption{Distributions of trajectory lengths (left), speed (middle) and acceleration (right) for GT (blue), RAW (orange) and REC (green) for scene iii (congested traffic).}
    \label{fig:dist_trajectories_ICCV_gt3}
\end{figure}

Next we show the distributions of trajectory lengths, velocities and accelerations for all three scenes in Figure~\ref{fig:dist_trajectories_ICCV_gt1}-\ref{fig:dist_trajectories_ICCV_gt3}. The data association step directly helps to produce longer trajectories, and the trajectory rectification step leads to a more realistic range of speed and acceleration, as compared to the raw tracking data which contains noises. 

% \usepackage{booktabs}

\begin{table*}[h]
\resizebox{\textwidth}{!}{%
\centering
\begin{tabular}{@{}l|llll|llll|llll@{}}
\toprule
Metrics                       & RAW-i & REC-i           &  &  & RAW-ii & REC-ii       &  &  & RAW-iii & REC-iii & & \\ 
\cmidrule(lr){3-5} \cmidrule(lr){7-9} \cmidrule(lr){11-13}
 & - & \textbf{Ours} & SD & Reversed & - & \textbf{Ours} & SD & Reversed  & - &\textbf{Ours} & SD & Reversed \\ \midrule
Precision $\uparrow$       & 0.71  & \textbf{0.90}& 0.58 &  0.87& 0.87 & \textbf{0.88} & 0.81 & 0.84 & 0.76 & \textbf{0.76} & 0.71 & 0.76\\
Recall $\uparrow$          & 0.56  & \textbf{0.83} & 0.79 &  0.81& 0.55 & \textbf{0.79}& 0.47 & 0.78& 0.46 & \textbf{0.67} & 0.47 & 0.65 \\
MOTA $\uparrow$            & 0.32  & \textbf{0.74}& 0.40 & 0.69 & 0.48 & \textbf{0.68 }& 0.36 & 0.65&  0.31& \textbf{0.46} & 0.28 & 0.44 \\
MOTP $\uparrow$            & 0.63  & \textbf{0.73} & 0.67 & 0.70 & 0.72 & \textbf{0.75}  & 0.73 &  0.74& 0.67 &\textbf{0.68} & 0.67 & 0.67\\
Fgmt/GT $\downarrow$ & 5.22  & \textbf{0.60}  &  1.97&4.28 & 5.38   & \textbf{1.93} & 3.55 & 2.65 & 4.92 & \textbf{1.11} & 3.32 & 5.28\\
Sw/GT $\downarrow$  & 1.43  & \textbf{0.04}  &  1.35& 0.11 & 2.98   & \textbf{0.53} &2.75  & 0.55 & 3.01 & \textbf{0.52} & 3.05 & 0.60\\ \bottomrule
\end{tabular}}
\caption{A comparison of data postprocessing procedures. Ours: data association with the matching cost specified in~\eqref{eq:cost}, followed by trajectory rectification; SD: data association with the simple distance (SD) metric as the matching cost, followed by trajectory rectification; Reversed: trajectory rectification first, then data association.}
\label{tab:compare_matching_costs}
\end{table*}
\edit{Lastly, we provide an additional experiment (Table~\ref{tab:compare_matching_costs}) to show the impact of the cost function $\Lambda(\phi_i, \phi_j)$ and data processing order on the reconstruction accuracy. Specifically, we compared our method with a simple distance (SD) metric as an alternative matching cost. The SD metric is specified as $\lVert p_i(t_e^i) - p_j(t_s^j)\rVert_2$, i.e., the Euclidean distance between the last measurement of fragment $i$ and the first measurement of fragment $j$. Additionally, we compared our 2-step procedure with the steps reversed, i.e., first run trajectory rectification followed by data association (``Reversed"). The results show that our original 2-step method with~\eqref{eq:cost} as the matching cost performs the best. }

In summary, the postprocessing improves all metrics on all the tested datasets. Particularly it produces longer trajectories with feasible dynamics. However, the most significant challenge occurs when associating fragments in dense and slow traffic, where the complex speed and lane-change behavior is not adequately captured by the current data association cost model.

% \begin{figure}
%     \centering
%     \includegraphics[width=\textwidth]{figures/TS_2022-12-02_resized.png}
%     \caption{Time-space diagram for four hours of I-24 WB morning rush hour traffic on Dec 2, 2022, generated from I-24 MOTION vehicle trajectories. x-axis: time of day (HH:MM); y-axis roadway postmile (mi).\Yanbing{replace wiht something slighlty different, make it taller}}
%     \label{fig:TS_2022-12-02}
% \end{figure}

% \begin{figure}
%     \centering
%     \includegraphics[width=\textwidth]{figures/postproc_architecture.pdf}
%     \caption{An overview of the postprocessing system diagram. The architecture includes several parallel processes performing local data association in each direction (EB: east bound, WB: west bound). The results are then passed on to the master processes, which perform association across adjacent road segments. The final step involves the reconciliation module, which imputes and smooths all associated fragments and writes the results to the database.}
%     \label{fig:postproc_architecture}
% \end{figure}

\subsection{\edit{Exp4: Scalability test on the complete I-24 MOTION system}}
\begin{figure}[h]
        \centering
        \includegraphics[width=\linewidth, trim={0 0 0 0.1cm},clip]{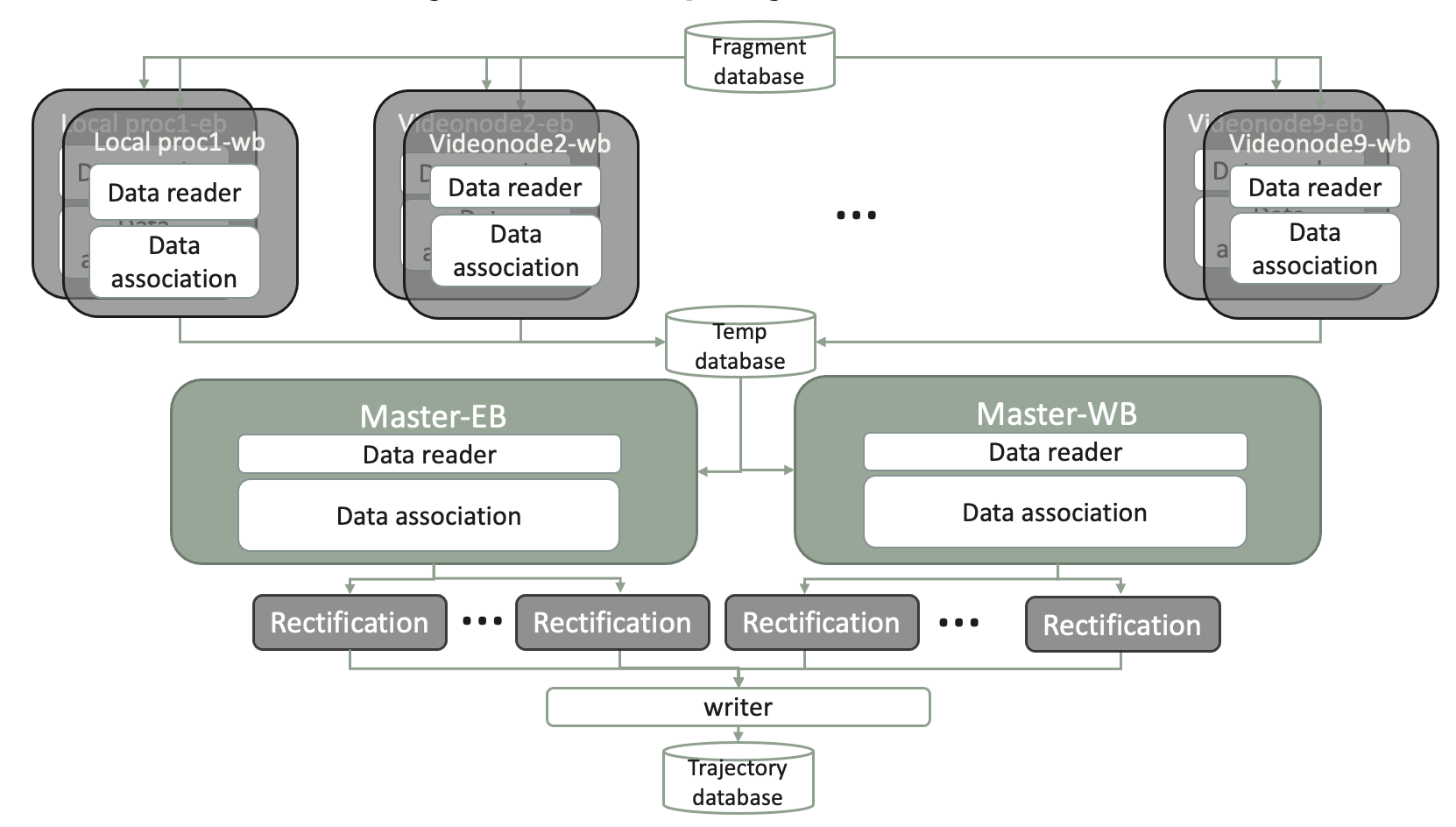}
        \caption{An overview of the postprocessing system diagram. The architecture includes several parallel processes performing local data association in each direction (EB: east bound, WB: west bound). The results are then passed on to the master processes, which perform association across adjacent road segments. The final step involves the trajectory rectification module, which imputes and smooths all associated fragments and writes the results to the database.}
        \label{fig:architecture}
    \end{figure}

The complete I-24 MOTION testbed~\cite{gloudemans202324} consists of 276 cameras in total that seamlessly cover 8 lanes (4.2 miles) of I-24 segment near Nashville, TN. On a typical workday morning, the recorded raw tracking data spans 4 hours, from 6:00 AM to 10:00 AM, making it 16 times larger in temporal scale, and 22 times larger in spatial scale than the NGSIM I-101 data. The dataset contains about 1.2 million tracklets (fragments). It is crucial to design an architecture that is scalable for size of the testbed.

To handle the volume of tracking data from this testbed, we design a postprocessing software architecture that employs parallel and asynchronous compute processes (see Figure~\ref{fig:architecture}). This architecture utilizes multiple processes running concurrently, managed by a master scheduler, to handle tracking data from nine upstream video processing nodes, with each process assigned to independently process fragments detected from 25-31 cameras. Each process performs local data association independently, and the locally processed results are passed to the master process of the corresponding direction of travel, which runs a second pass of data association to connect the partial trajectory fragments across each adjacent local process. Finally, all resulting trajectories are smoothed and imputed in the ``reconciliation module" before being written to the postprocessed database. 

Our experiments demonstrate that this software architecture can effectively handle the volume of tracking data from this testbed. In addition to implementing an online version of the negative cycle cancellation algorithm mentioned in Section~\ref{sec:online_ncc}, our approach can process approximately 400 trajectories per minute for light traffic (flow is about 30 vehicles/lane/min), while for heavy traffic (120 vehicles/lane/min) it can process one minute of data in 50 seconds. For a full-scale 4-hour run with the asynchronous software structure illustrated in Figure~\ref{fig:architecture}, local processes are completed in 1.5 hours, and master processes in 2 hours, resulting in a total runtime of 3.5 hours to process 4 hours of data. The server that hosts this architecture operates on Ubuntu 20.04, with 528GB total memory and 64 CPU cores. These results indicate that our proposed algorithm is scalable and can efficiently process large volumes of real-time tracking data on I-24 MOTION.

\section{Conclusion and future work}
\label{sec:conclusions}
High-quality trajectory data can close the gap for understanding microscopic traffic phenomena. A real-world live testbed like I-24 MOTION helps researchers to understand the impact of mixed autonomy in traffic. However, data produced by cameras and upstream computer vision algorithms still lacks high quality to be ready for research use. In this paper we demonstrate a two-step data postprocessing pipeline to automatically reconcile detection and tracking data. The pipeline includes a fragment association algorithm to solve an online min-cost flow problem, and a trajectory rectification approach formulated as a quadratic programming. The accuracy is benchmarked on both numerical experiments using miscrosimulation, NGSIM data, as well as on the raw tracking data from three scenes of the I-24 MOTION system and the corresponding manually labeled ground truth. Results show that the two-step treatments improve a variety of trajectory quality measures on all the testing cases, given different traffic scenarios. Noticeably, it significantly improves the velocity and acceleration dynamics, despite the parameter tuning and cost function design step are only preliminary. This proposed pipeline has high promises to replace previous manual efforts on data cleaning. Additionally, we show a design of a software architecture for postprocessing that utilizes asynchronous processes and is currently deployed on the I-24 MOTION system to continuously generate a large volume of high-quality open-road trajectory data. 

For future work, a few algorithmic improvements can be made on existing methods to potentially address issues in real data. For example, a multi-vehicle reconciliation formulation can be considered to account for potential collisions. Furthermore, different cost models for data association and an exhaustive parameter tuning need to be performed to improve the current results.

\section*{Acknowledgement}
This study is based upon work supported by the National Science Foundation (NSF) under Grant No. 2135579, the NSF Graduate Research Fellowship Grant No. DGE-1937963 and the USDOT Dwight D. Eisenhower Fellowship program under Grant No. 693JJ32245006. The authors are grateful to Caliper for technical support on the TransModeler micro-simulation software used in this work.

%%       \citet{<label>} ==> Jones et al. (2015)
%%       \citep{<label>} ==> (Jones et al., 2015)

\newpage
\appendix
\section{Synthetic data}
\label{app:tm_data}
\begin{figure}[ht]
    \centering
    \includegraphics[width=0.9\linewidth]{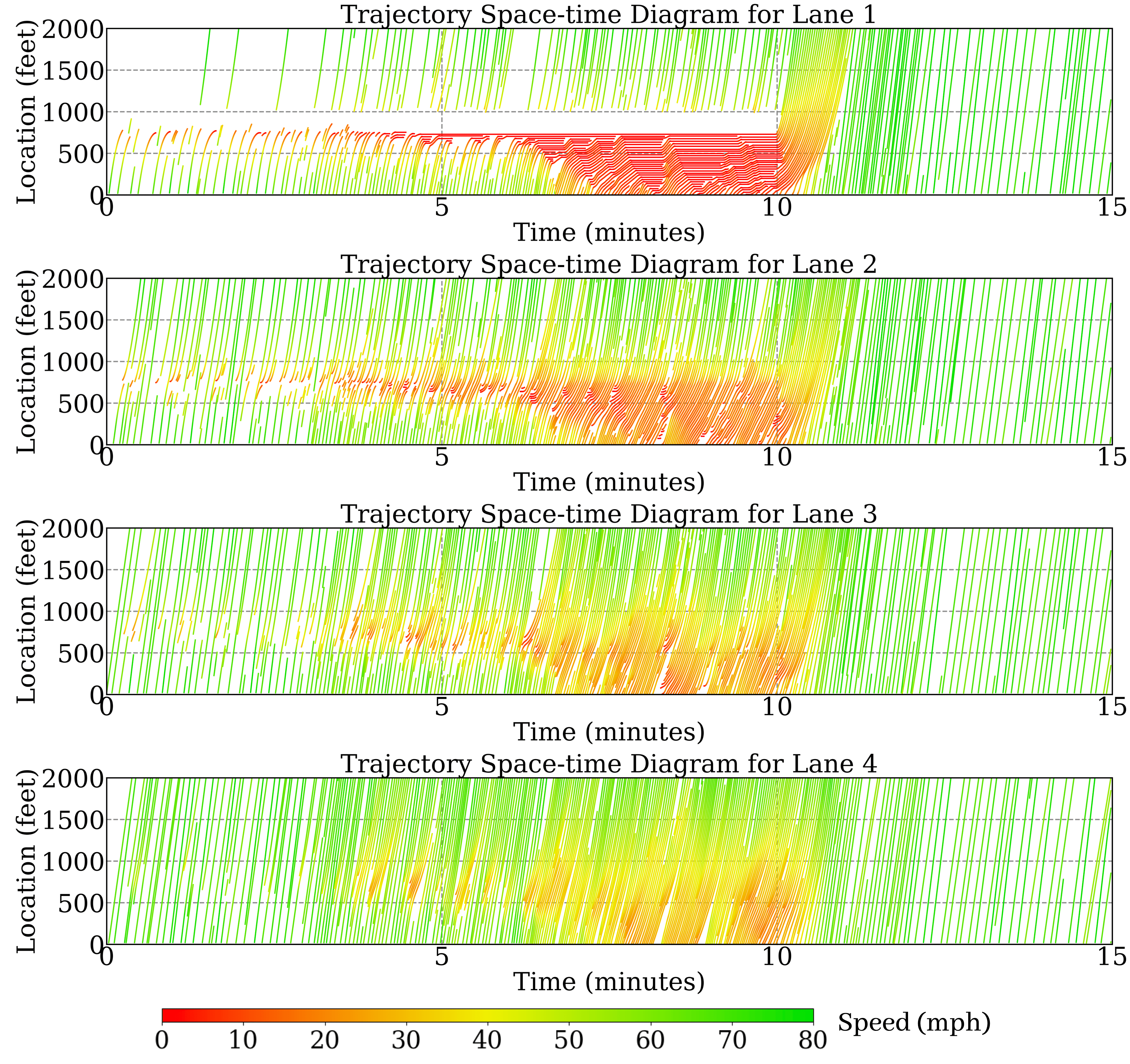}
    \caption{Trajectory space-time diagram for the synthetic data (SIM 3): figures from top to bottom correspond to traffic in lane 1-4, respectively.}
    \label{fig:gt}
\end{figure}

The synthetic data used to test the postprocessing pipeline was generated using TransModeler 6.1, a micro-simulation software. The simulation involved a one-way, 2000ft-long four-lane highway, with a capacity of 2000 $veh/hr/lane$, and lasted 15 minutes (900 seconds). To evaluate the algorithms' performance under varying traffic conditions, the simulation incorporated time-varying traffic demand and a bottleneck. The traffic demand varied between 1200 $veh/hr/lane$ in 1-3 minutes and 13-15 minutes, 2400 $veh/hr/lane$ in 4-6 minutes and 10-12 minutes, and 3600 $veh/hr/lane$ in 6-9 minutes. The bottleneck was triggered by a lane closure signal, with lane-1 (leftmost) closed for the first 10 minutes. Figure \ref{fig:gt} depicts the trajectory space-time diagram for the synthetic data. The lane-changing logic in TransModeler involves a decision process with multiple rules, including the target lane rule and the gap acceptance rule \citep{yang1996microscopic}. The target lane rule determines which lane to change to, while the gap acceptance rule determines whether the gap on the targeted lane is acceptable to the vehicle. The lane-changing action is triggered when both rules are satisfied. It should be noted that the car-following and lane-changing model parameters were set to default values in TransModeler.

\begin{figure}[ht]
    \centering
    \includegraphics[width=1.0\linewidth]{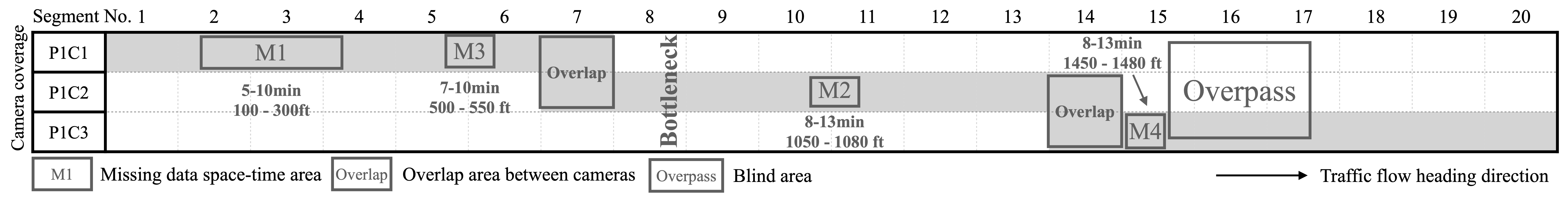}
    \caption{A segment with multiple cameras reproduced in micro-simulation}
    \label{fig:ppid}
\end{figure}

In order to simulate the traffic observation system that utilizes multiple cameras, the 2000ft-long highway sketch is divided into 20 segments, each 100ft long (see Figure~\ref{fig:ppid}). The highway is monitored by three cameras, namely P1C1, P1C2, and P1C3, with P1C1 covering segments 1 to 7, P1C2 covering segments 7 to 14, and P1C3 covering segments 14 to 20. Each camera assigns a different ID for the trajectory of the same vehicle, and there is an overlap area between adjacent cameras, such as segment 7 for P1C1 and P1C2, and segment 14 for P1C2 and P1C3. To replicate real-world scenarios, typical data loss situations such as overpass and camera packet loss (M1, M2, M3) are simulated. For instance, M1 (5-10min, 100-300ft) indicates the data in the space-time zone from 5 to 10 minutes at 100ft to 300 ft is lost. As in real-world tracking tasks, when a vehicle passes through a missing data area, its ID will be changed and it will be re-initialized as a new object in the system. The overpass area mechanism is similar.
% \Junyi{Maybe we could share the synthetic data to the reviewers.}

% \Junyi{I need to unify the name used in Figure \ref{fig:ppid}, such as overpasses, misaligned camera field of views, and camera package loss. Check it again before submission.}


% \bibliographystyle{elsarticle-harv} 
\bibliographystyle{abbrv}
\newpage
\bibliography{reference}





% \end{thebibliography}
\end{document}